\documentclass[aip,jcp,amsmath,amssymb,preprint]{revtex4-1}

\usepackage{dcolumn} 
\usepackage{bm}
\usepackage{graphicx}
\usepackage{longtable}
\usepackage{footnote}
\usepackage{float}

\usepackage{color}

\draft 

\begin{document}

\newcommand{\bra}[1]{\langle #1|}
\newcommand{\ket}[1]{|#1\rangle}
\newcommand{\braket}[2]{\langle #1|#2\rangle}
\newcommand{\p}{^\prime}
\newcommand{\pp}{^{\prime\prime}}

\title{A global potential energy surface and dipole moment surface for silane}

\author{Alec Owens} 
 \email{owens@mpi-muelheim.mpg.de}
\affiliation{Max-Planck-Institut f\"{u}r Kohlenforschung, Kaiser-Wilhelm-Platz 1, 45470 M\"{u}lheim an der Ruhr, Germany}
\affiliation{Department of Physics and Astronomy, University College London, Gower Street, WC1E 6BT London, United Kingdom}

\author{Sergei N. Yurchenko}
\affiliation{Department of Physics and Astronomy, University College London, Gower Street, WC1E 6BT London, United Kingdom}

\author{Andrey Yachmenev}
\affiliation{Department of Physics and Astronomy, University College London, Gower Street, WC1E 6BT London, United Kingdom}

\author{Walter Thiel}
\affiliation{Max-Planck-Institut f\"{u}r Kohlenforschung, Kaiser-Wilhelm-Platz 1, 45470 M\"{u}lheim an der Ruhr, Germany}

\date{\today}

\begin{abstract}
 A new nine-dimensional potential energy surface (PES) and dipole moment surface (DMS) for silane have been generated using high-level \textit{ab initio} theory. The PES, CBS-F12$^{\,\mathrm{HL}}$, reproduces all four fundamental term values for $^{28}$SiH$_4$ with sub-wavenumber accuracy, resulting in an overall root-mean-square (rms) error of $0.63{\,}$cm$^{-1}$. The PES is based on explicitly correlated coupled cluster calculations with extrapolation to the complete basis set limit, and incorporates a range of higher-level additive energy corrections to account for core-valence electron correlation, higher-order coupled cluster terms, and scalar relativistic effects. Systematic errors in computed intra-band rotational energy levels are reduced by empirically refining the equilibrium geometry. The resultant Si-H bond length is in excellent agreement with previous experimental and theoretical values. Vibrational transition moments, absolute line intensities of the $\nu_3$ band, and the infrared spectrum for $^{28}$SiH$_4$ including states up to $J=20$ and vibrational band origins up to $5000{\,}$cm$^{-1}$ are calculated and compared with available experimental results. The DMS tends to marginally overestimate the strength of line intensities. Despite this, band shape and structure across the spectrum are well reproduced and show good agreement with experiment. We thus recommend the PES and DMS for future use.
\end{abstract}

\pacs{}

\maketitle 

\section{Introduction}
\label{sec:intro}
 
 The infrared (IR) absorption spectrum of silane (SiH$_4$) was first documented over eighty years ago.~\cite{34StNixx.SiH4,35StNixx.SiH4} Since then numerous high-resolution spectroscopic studies of SiH$_4$ and its isotopomers have followed, including astronomical observation of rotation-vibration transitions around the carbon star IRC +10216~\cite{84GoBexx.SiH4,93KeRixx.SiH4,00MoDaHa.SiH4} and in the atmospheres of Jupiter~\cite{78TrLaFi.SiH4} and Saturn.~\cite{80LaFiSm.SiH4} In industry silane gas is used extensively in the semiconductor manufacturing process and for the production of solar cells.
 
 Despite its industrial and astrophysical importance, very few rigorous theoretical studies have been carried out. \citet{99MaBaLe.SiH4} computed an accurate quartic force field for silane based on CCSD(T) [coupled cluster with all single and double excitations and a perturbational estimate of connected triple excitations] calculations using the correlation consistent quadruple zeta basis set, cc-pVQZ,~\cite{Dunning89} plus an additional high-exponent \textit{d}-function~\cite{Martin:1998} (denoted as cc-pVQZ+1 in Ref.~\onlinecite{99MaBaLe.SiH4}). Minor empirical refinement of the four diagonal quadratic constants produced a force field of spectroscopic quality ($\pm 1{\,}$cm$^{-1}$ when reproducing the fundamental frequencies) applicable for several isotopomers of silane.
 
 The resultant force field was subsequently used to calculate vibrational energy levels of SiH$_4$, SiH$_3$D, SiHD$_3$, and SiH$_2$D$_2$ by means of canonical Van-Vleck perturbation theory (CVPT).~\cite{00WaSixx.SiH4} When compared to results of a variational four-dimensional stretch model, full-dimensional CVPT calculations were necessary to accurately describe certain stretch levels as they incorporated the effects of Fermi resonance. The importance of treating Fermi interactions to compute vibrational energies of silane was also highlighted previously using an algebraic approach.~\cite{01HoBoBe.SiH4} 
 
 The use of stretch-only models has generally been successful in describing stretching overtones~\cite{82HaChxx.SiH4,88HaChxx.SiH4,97PeCaxx.SiH4,02XiTexx.SiH4} and corresponding band intensities~\cite{88HaChxx.SiH4,98LiWaCh.SiH4,99LiYuZh.SiH4,01LiHeWa.SiH4,02HeLiLi.SiH4}  however. This is because of the pronounced local mode behaviour of silane, the effects of which have been documented experimentally in a series of papers by Zhu et al.~\cite{89ZhZhMa.SiH4,90ZhZhMa.SiH4,90ZhZhMa.SiH4,90ZhMaZh.SiH4,91ZhQiMa.SiH4,95SuWaZh.SiH4} It is only at higher energies (above $12{\,}000\,$cm$^{-1}$) that the rotational structure of the $\ket{6000}$ and $\ket{7000}$ stretch eigenstates can no longer be analysed in a local mode description due to vibrational resonances.~\cite{94ZhCaSt.SiH4} For intensity calculations, even a small treatment of bending motion can improve the description of intensities compared to stretch-only models~\cite{01LiBuHe.SiH4} (an overview of previously computed \textit{ab initio} dipole moment surfaces for silane can be found in Ref.~\onlinecite{13Yuxxxx.method}).

 The motivation for the present work is that $^{28}$SiH$_4$ (henceforth labelled as SiH$_4$) is a target molecule of the ExoMol project,~\cite{ExoMol2012} which is creating a comprehensive database of all molecular transitions deemed necessary to model exoplanet and other hot atmospheres. Although unlikely, SiH$_4$ has already been considered in the context of biosignature gases on rocky exoplanets.~\cite{13aSeBaHu.CH3Cl}
 
 At present there is no coverage of SiH$_4$ in several of the popular spectroscopic databases.~\cite{HITRAN,GEISA,JPL,CDMS:2005} The PNNL spectral library~\cite{PNNL} is an exception, covering the range of $600$ to $6500\,$cm$^{-1}$ at a resolution of around $0.06\,$cm$^{-1}$ for temperatures of 5, 25, and 50$\,^{\circ}$C. The Spherical Top Data System~\cite{STDS:1998} (STDS) is another valuable resource for spectral information on silane. However, some of the measured transitions and intensities are from unpublished work which makes it hard to verify the methods used and subsequently the reliability of the data.

 It is our intention to construct a global nine-dimensional potential energy surface (PES) and dipole moment surface (DMS) for silane. To do this we employ state-of-the-art electronic structure calculations to generate the respective surfaces. After fitting the \textit{ab initio} data with suitable analytic representations, the quality of the PES and DMS will be tested by means of variational calculations of the infrared spectrum. 

  The paper is structured as follows: In Sec.~\ref{sec:PES} the \textit{ab initio} calculations and analytic representation of the PES are presented. Similarly, in Sec.~\ref{sec:DMS} the electronic structure calculations and analytic representation of the DMS are detailed. Pure rotational energies, the equilibrium Si-H bond length, vibrational $J=0$ energy levels, absolute line intensities of the $\nu_3$ band, and an overview of the rovibration spectrum up to $J=20$ are calculated and compared against available experimental data in Sec.~\ref{sec:results}. We offer concluding remarks in Sec.~\ref{sec:conc}.
 
\section{Potential Energy Surface}
\label{sec:PES} 

\subsection{Electronic structure calculations}
 
 Focal-point analysis~\cite{Csaszar98} is used to represent the total electronic energy as
\begin{equation}\label{eq:tot_en}
E_{\mathrm{tot}} = E_{\mathrm{CBS}}+\Delta E_{\mathrm{SR}}+\Delta E_{\mathrm{CV}}+\Delta E_{\mathrm{HO}}
\end{equation}

\noindent The energy at the complete basis set (CBS) limit $E_{\mathrm{CBS}}$ was computed using the explicitly correlated F12 coupled cluster method CCSD(T)-F12b~(Ref.~\onlinecite{Adler07}) with the F12-optimized correlation consistent polarized valence basis sets, cc-pVTZ-F12 and cc-pVQZ-F12.~\cite{Peterson08} Calculations were carried out in the frozen core approximation and used the diagonal fixed amplitude ansatz 3C(FIX)~\cite{TenNo04} with a Slater geminal exponent value of $\beta=1.0\,a_0^{-1}$.~\cite{Hill09} For the resolution of the identity (RI) basis and the two density fitting (DF) basis sets, we employed the corresponding OptRI,~\cite{Yousaf08} cc-pV5Z/JKFIT,~\cite{Weigend02} and aug-cc-pwCV5Z/MP2FIT~\cite{Hattig05} auxiliary basis sets (ABS), respectively. All calculations were carried out with MOLPRO2012~\cite{Werner2012} unless stated otherwise.
 
 A parameterized two-point formula, $E^{C}_{\mathrm{CBS}} = (E_{n+1} - E_{n})F^{C}_{n+1} + E_{n}$, proposed by Hill et al.~\cite{Hill09} was used to extrapolate to the CBS limit. For the coefficients $F^{C}_{n+1}$, which are specific to the CCSD-F12b and (T) components of the total CCSD(T)-F12b energy, we employed values of $F^{\mathrm{CCSD-F12b}}=1.363388$ and $F^{\mathrm{(T)}}=1.769474$ as recommended in Ref.~\onlinecite{Hill09}. The Hartree-Fock (HF) energy was not extrapolated. Instead the HF+CABS (complementary auxiliary basis set) singles correction~\cite{Adler07} calculated in the larger basis set was used.

 The scalar relativistic (SR) correction $\Delta E_{\mathrm{SR}}$ was computed using the second-order Douglas-Kroll-Hess approach~\cite{dk1,dk2} at the CCSD(T)/cc-pVQZ-DK~\cite{dk_basis} level of theory in the frozen core approximation. The spin-orbit interaction was not considered as for light, closed-shell molecules it can be safely ignored in spectroscopic calculations.~\cite{Tarczay01}
 
 The core-valence (CV) electron correlation correction $\Delta E_{\mathrm{CV}}$ was calculated at the CCSD(T)-F12b level of theory in conjunction with the F12-optimized correlation consistent core-valence basis set cc-pCVTZ-F12.~\cite{Hill10} The same ansatz and ABS as in the frozen core approximation computations were used, however we set $\beta=1.4\,a_0^{-1}$. The (1\textit{s}) orbital of Si was frozen for all-electron calculations.
 
 To estimate the higher-order (HO) correction $\Delta E_{\mathrm{HO}}$ we used the hierarchy of coupled cluster methods such that $\Delta E_{\mathrm{HO}} = \Delta E_{\mathrm{T}} + \Delta E_{\mathrm{(Q)}}$. Here the full triples contribution is $\Delta E_{\mathrm{T}} = \left[E_{\mathrm{CCSDT}}-E_{\mathrm{CCSD(T)}}\right]$, and the perturbative quadruples contribution is $\Delta E_{\mathrm{(Q)}} = \left[E_{\mathrm{CCSDT(Q)}}-E_{\mathrm{CCSDT}}\right]$. Calculations were carried out in the frozen core approximation at the CCSD(T), CCSDT, and CCSDT(Q) levels of theory using the general coupled cluster approach~\cite{Kallay05,Kallay08} as implemented in the MRCC code~\cite{mrcc} interfaced to CFOUR.~\cite{cfour} The full triples computation utilized the correlation consistent triple zeta basis set, cc-pVTZ(+d for Si),~\cite{Dunning89,Kendall92,Woon93,Dunning01} whilst the perturbative quadruples computation employed the double zeta basis set, cc-pVDZ(+d for Si).
 
 The contribution from the diagonal Born-Oppenheimer correction (DBOC) was computed with all electrons correlated (bar the (1\textit{s}) orbital of Si) using the CCSD method~\cite{Gauss06} as implemented in CFOUR with the aug-cc-pCVDZ basis set. A preliminary analysis of the DBOC on the vibrational energy levels showed no improvement overall when compared against experimental values. Given that inclusion of the DBOC means the PES becomes applicable only for $^{28}$SiH$_4$ and no other isotopologues, the correction was not included.

\begin{figure*}
\includegraphics[width=\textwidth,angle=0]{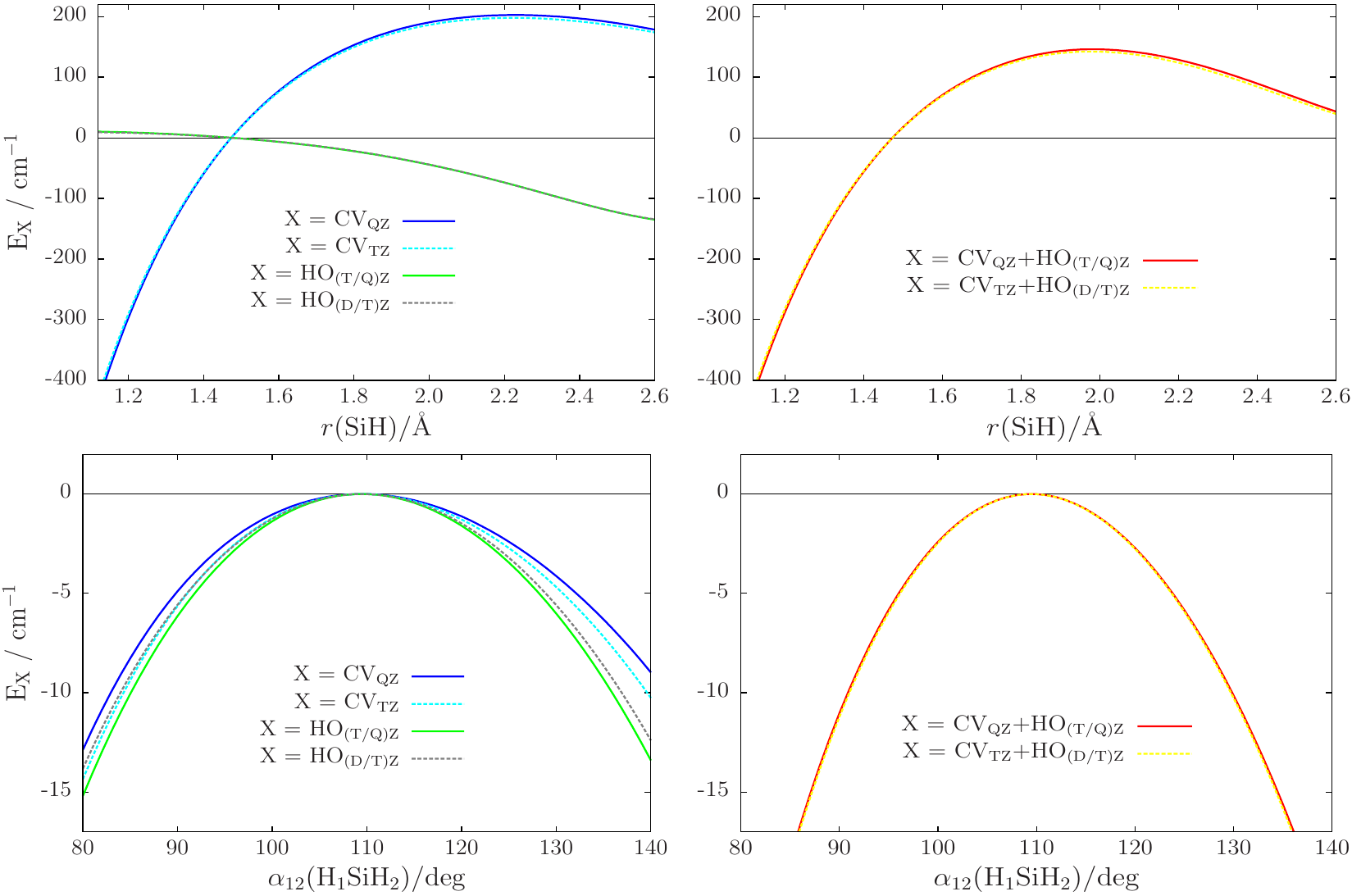}
\caption{\label{fig:1d:cv_ho}One-dimensional cuts of the CV, HO, and CV+HO corrections for different sizes of basis set. For CV the subscript TZ(QZ) refers to calculations with the cc-pCVTZ-F12(cc-pCVQZ-F12) basis set. For HO the subscript (D/T)Z refers to calculations with the cc-pVDZ and cc-pVTZ basis sets for the perturbative quadruples and full triples, respectively. Likewise the (T/Q)Z subscript corresponds to the cc-pVTZ and cc-pVQZ basis sets.}
\end{figure*}

 In generating a high-level \textit{ab initio} PES for silane we have opted for a more pragmatic approach. Obtaining tightly converged energies with respect to basis set size for the HL corrections is less important, particularly for the CV and HO contributions which are computationally more demanding. Since the CV and HO corrections usually enter the electronic energy with opposing sign, we have calculated them together utilizing smaller basis sets. Although independently the separate corrections are not fully converged, this error is compensated for when considering their sum. This is illustrated through one-dimensional cuts of the PES in Fig.~\eqref{fig:1d:cv_ho}, most noticeably in the bending cut.
 
 The global grid was built in terms of nine internal coordinates; four Si-H bond lengths $r_1$, $r_2$, $r_3$, $r_4$, and five $\angle(\mathrm{H}_j$-Si-$\mathrm{H}_k)$ interbond angles $\alpha_{12}$, $\alpha_{13}$, $\alpha_{14}$, $\alpha_{23}$, and $\alpha_{24}$, where $j$ and $k$ label the respective hydrogen atoms. The Si-H stretch distances ranged from $0.98\leq r_i \leq 2.95{\,}\mathrm{\AA}$ for $i=1,2,3,4$ whilst bending angles varied from $40\leq \alpha_{jk} \leq 140^{\circ}$ where $jk=12,13,14,23,24$. All terms in Eq.~\eqref{eq:tot_en} were calculated on a grid of $84{\,}002$ geometries with energies up to $h c \cdot 50{\,}000{\,}$cm$^{-1}$ ($h$ is the Planck constant and $c$ is the speed of light). At every grid point the coupled cluster energies were extrapolated to the CBS limit, and each HL correction was calculated and added to the total electronic energy.
 
 The HL corrections have been computed at each grid point which is in fact time-effective at the levels of theory chosen for the electronic structure calculations. The alternative is to design reduced grids for each correction, fit a corresponding analytic representation and apply the resulting form to the global grid of geometries by interpolation (see Refs.~\onlinecite{YaYuRi11.H2CS,15OwYuYa.CH3Cl} for examples of this strategy). Although this alternative is computationally less intensive, achieving a satisfactory description of each HL correction requires careful consideration and may not be trivial; any such problems are avoided in our present approach.
 
\subsection{Analytic representation}

 The analytic representation chosen for the present study has previously been used for methane.~\cite{13YuTeBa.CH4,14YuTexx.CH4,14YuTeBa.CH4} For the stretch coordinates,
\begin{equation}\label{eq:stretch}
\xi_i=1-\exp\left(-a(r_i - r^{\mathrm{ref}})\right){\,};\hspace{2mm}i=1,2,3,4
\end{equation}

\noindent where $a=1.47{\,}\mathrm{\AA}^{-1}$ and the reference equilibrium structural parameter $r^{\mathrm{ref}}=1.4741{\,}\mathrm{\AA}$ (value discussed in Sec.~\ref{sec:results}). The angular terms are given as symmetrized combinations of interbond angles,
\begin{equation}\label{eq:ang1}
\xi_5 = \frac{1}{\sqrt{12}}\left(2\alpha_{12}-\alpha_{13}-\alpha_{14}-\alpha_{23}-\alpha_{24}+2\alpha_{34}\right)
\end{equation}
\begin{equation}\label{eq:ang2}
\xi_6 = \frac{1}{2}\left(\alpha_{13}-\alpha_{14}-\alpha_{23}+\alpha_{24}\right)
\end{equation}
\begin{equation}\label{eq:ang3}
\xi_7 = \frac{1}{\sqrt{2}}\left(\alpha_{24}-\alpha_{13}\right)
\end{equation}
\begin{equation}\label{eq:ang4}
\xi_8 = \frac{1}{\sqrt{2}}\left(\alpha_{23}-\alpha_{14}\right)
\end{equation}
\begin{equation}\label{eq:ang5}
\xi_9 = \frac{1}{\sqrt{2}}\left(\alpha_{34}-\alpha_{12}\right)
\end{equation}

\noindent The potential function (maximum expansion order of $i+j+k+l+m+n+p+q+r=6$),
\begin{equation}\label{eq:pot_f}
V(\xi_{1},\xi_{2},\xi_{3},\xi_{4},\xi_{5},\xi_{6},\xi_{7},\xi_{8},\xi_{9})={\sum_{ijk\ldots}}{\,}\mathrm{f}_{ijk\ldots}V_{ijk\ldots}
\end{equation}

\noindent contains the terms
\begin{equation}
V_{ijk\ldots}=\lbrace\xi_{1}^{\,i}\xi_{2}^{\,j}\xi_{3}^{\,k}\xi_{4}^{\,l}\xi_{5}^{\,m}\xi_{6}^{\,n}\xi_{7}^{\,p}\xi_{8}^{\,q}\xi_{9}^{\,r}\rbrace^{\bm{T}_{\mathrm{d}}\mathrm{(M)}}
\end{equation}

\noindent which are symmetrized combinations of different permutations of the coordinates $\xi_{i}$, and transform according to the $\bm{T}_{\mathrm{d}}\mathrm{(M)}$ molecular symmetry group.~\cite{MolSym_BuJe98} They are found by solving an over-determined system of linear equations in terms of the nine coordinates given above. A total of 287 symmetrically unique terms were derived up to sixth order of which only 104 were employed for the final PES. The corresponding expansion parameters $\mathrm{f}_{ijk\ldots}$ were determined from a least-squares fitting to the \textit{ab initio} data. Weight factors of the form,~\cite{Schwenke97}
\begin{equation}\label{eq:weights}
w_i=\left(\frac{\tanh\left[-0.0006\times(\tilde{E}_i - 15{\,}000)\right]+1.002002002}{2.002002002}\right)\times\frac{1}{N\tilde{E}_i^{(w)}}
\end{equation}

\noindent were used in the fit. Here $\tilde{E}_i^{(w)}=\max(\tilde{E}_i, 10{\,}000)$, where $\tilde{E}_i$ is the potential energy at the $i$th geometry above equilibrium and the normalization constant $N=0.0001$ (all values in cm$^{-1}$). The final fitted PES required 106 expansion parameters and employed Watson's robust fitting scheme,~\cite{Watson03} which reduces the weights of outliers and improves the fit at lower energies. A weighted root-mean-square (rms) error of $1.77{\,}$cm$^{-1}$ was obtained for energies up to $h c \cdot 50{\,}000{\,}$cm$^{-1}$.

  Note that geometries with $r_i\geq 2.30{\,}\mathrm{\AA}$ for $i=1,2,3,4$ possessed a T1 diagnostic value $>0.02$,~\cite{T1_Lee89} and so the corresponding weights were reduced by several orders of magnitude. Although the coupled cluster method is not completely accurate at these points, by including them the PES maintains a reasonable shape towards dissociation. In subsequent calculations we refer to this PES as CBS-F12$^{\,\mathrm{HL}}$. The CBS-F12$^{\,\mathrm{HL}}$ expansion parameter set is provided in the supplementary material along with a FORTRAN routine to construct the PES.~\cite{EPAPSSIH4}
  
\section{Dipole Moment Surface}
\label{sec:DMS} 

\subsection{Electronic structure calculations}

  The electric dipole moment is equal to the first derivative of the electronic energy with respect to external electric field strength. For each of the $X$, $Y$, and $Z$ Cartesian coordinate axes with origin at the Si nucleus, an external electric field with components $\pm0.005{\,}$a.u. was applied and the dipole moment components $\mu_X$, $\mu_Y$, and $\mu_Z$ computed by means of the central finite difference scheme. Calculations were carried out at the CCSD(T)/aug-cc-pVTZ(+d for Si) level of theory in the frozen core approximation using MOLPRO2012. The same nine-dimensional grid as used for the PES with energies up to $h c \cdot 50{\,}000{\,}$cm$^{-1}$ was employed.

\subsection{Analytic representation}

  To represent the dipole moment surface (DMS) analytically it is necessary to transform to a suitable molecule-fixed $xyz$ coordinate system. For the present study we utilize the symmetrized molecular bond (SMB) representation for XY$_4$ molecules.~\cite{13YuTeBa.CH4} We first define unit vectors along the four Si-H bonds,
\begin{equation}
\mathbf{e}_i = \frac{\mathbf{r}_i-\mathbf{r}_0}{\lvert \mathbf{r}_i-\mathbf{r}_0\rvert}{\,};\hspace{4mm}i=1,2,3,4
\end{equation}

\noindent where $\mathbf{r}_0$ is the position vector of the Si nucleus, and $\mathbf{r}_i$ is that of the respective H$_i$ atom. Three symmetrically independent reference vectors which span the $F_2$ representation are formed,
\begin{equation}
\mathbf{n}_1 = \frac{1}{2}\left(\mathbf{e}_1-\mathbf{e}_2+\mathbf{e}_3-\mathbf{e}_4\right)
\end{equation}
\begin{equation}
\mathbf{n}_2 = \frac{1}{2}\left(\mathbf{e}_1-\mathbf{e}_2-\mathbf{e}_3+\mathbf{e}_4\right)
\end{equation}
\begin{equation}
\mathbf{n}_3 = \frac{1}{2}\left(\mathbf{e}_1+\mathbf{e}_2-\mathbf{e}_3-\mathbf{e}_4\right)
\end{equation}

\noindent Using these the \textit{ab initio} dipole moment vector $\bm{\mu}$ can be expressed as
\begin{equation}
\bm{\mu} = \mu_x\mathbf{n}_1+\mu_y\mathbf{n}_2+\mu_z\mathbf{n}_3
\end{equation}

\noindent Here $\mu_{\alpha}$ ($\alpha=x,y,z$) are the dipole moment functions (also of $F_2$ symmetry) which take the form
\begin{equation}\label{eq:mu_tot}
\mu_{\alpha}(\xi_{1},\xi_{2},\xi_{3},\xi_{4},\xi_{5},\xi_{6},\xi_{7},\xi_{8},\xi_{9})={\sum_{ijk\ldots}}F^{(\alpha)}_{ijk\ldots}\mu^{F_2}_{\alpha,ijk\ldots}
\end{equation}

\noindent The expansion terms 
\begin{equation}
\mu^{F_2}_{\alpha,ijk\ldots}=\lbrace\xi_{1}^{\,i}\xi_{2}^{\,j}\xi_{3}^{\,k}\xi_{4}^{\,l}\xi_{5}^{\,m}\xi_{6}^{\,n}\xi_{7}^{\,p}\xi_{8}^{\,q}\xi_{9}^{\,r}\rbrace^{F_{2\alpha}}
\end{equation}

\noindent are symmetrized combinations of different permutations of coordinates $\xi_{i}$, and span the $F_{2\alpha}$ representation of the $\bm{T}_{\mathrm{d}}\mathrm{(M)}$ molecular symmetry group (see Ref.~\onlinecite{13YuTeBa.CH4} for more detail). A sixth order expansion was employed in terms of the coordinates,
\begin{equation}
\xi_i=(r_i-r^{\mathrm{ref}})\exp\left(-\beta(r_i-r^{\mathrm{ref}})^{2}\right){\,};\hspace{2mm}i=1,2,3,4
\end{equation}

\noindent for the stretches, with the same angular coordinates as before (Eqs.~\eqref{eq:ang1} to \eqref{eq:ang5}). The factor $\exp\left(-\beta(r_i-r^{\mathrm{ref}})^{2}\right)$  prevents the expansion from diverging at large values of $r_i$. Our DMS fitting employed the parameters $r^{\mathrm{ref}}=1.5355{\,}\mathrm{\AA}$ and $\beta=1.0{\,}\mathrm{\AA}^{-2}$. 

 The expansion coefficients $F^{(\alpha)}_{ijk\ldots}$ for all three components $\alpha=x,y,z$ were determined simultaneously through a least squares fitting to the \textit{ab initio} data. Again weight factors of the form given in Eq.~\eqref{eq:weights} were used which favor energies below $h c \cdot 15{\,}000{\,}$cm$^{-1}$. The fitting required 283 parameters and reproduced the \textit{ab initio} data with a weighted rms error of $0.001{\,}$D for energies up to $h c \cdot 50{\,}000{\,}$cm$^{-1}$. The expansion parameter set for the DMS is provided in the supplementary material along with a FORTRAN routine to construct the corresponding analytic representation.~\cite{EPAPSSIH4}
     
\section{Results}
\label{sec:results}

\subsection{Equilibrium bond length and pure rotational energies}
\label{sec:eq_rotational}

 Since rotational energies are highly dependent on the molecular geometry through the moments of inertia, we first refine the Si-H reference equilibrium structural parameter $r^{\mathrm{ref}}$ before we proceed to extensive rovibrational energy level calculations. Thereby, the accuracy of the computed intra-band rotational wavenumbers can be significantly improved.~\cite{YuBaYa09.NH3,13YaPoTh.H2CS} 
 
 Two iterations of a nonlinear least-squares fit to the experimental $J\leq 6$ rotational energies from Ref.~\onlinecite{STDS:1998} produced a refined parameter of $r^{\mathrm{ref}}=1.4741{\,}\mathrm{\AA}$. However, due to the inclusion of a linear expansion term in the parameter set of our potential, this value does not define the minimum of the PES. The true equilibrium bond length was determined to be $r^{\mathrm{eq}}=1.4737{\,}\mathrm{\AA}$. This is in good agreement with the experimental estimate of $r\mathrm{(Si\!-\!H)}=1.4741{\,}\mathrm{\AA}$,~\cite{85OhMaEn.SiH4} and an \textit{ab initio} value of $r\mathrm{(Si\!-\!H)}=1.4742{\,}\mathrm{\AA}$ calculated at the all electron CCSD(T)/cc-pCVQZ level of theory.~\citep{05CoMaGa.SiH4} Note that before the refinement the original \textit{ab initio} bond length of the CBS-F12$^{\,\mathrm{HL}}$ PES was $r^{\mathrm{eq}}_{\mathit{ab\,initio}}=1.4735{\,}\mathrm{\AA}$.
 
 The computed pure rotational energies are listed in Table~\ref{tab:rotational}. The details of the calculations will be discussed in Sec.~\ref{sec:J0_energies}. As can be seen, the agreement with experiment is excellent and energy levels up to $J\leq 6$ are reproduced with a rms error of $0.00005\,$cm$^{-1}$. We therefore expect the true Si-H equilibrium bond length to be very close to the value $r^{\mathrm{eq}}=1.4737{\,}\mathrm{\AA}$.

\begin{table}[!h]
\tabcolsep=5pt
\caption{\label{tab:rotational}Comparison of calculated and experimental $J\leq6$ pure rotational term values (in cm$^{-1}$) for $^{28}$SiH$_4$. The observed ground state energy levels are from Ref.~\onlinecite{STDS:1998}.}
\begin{center}
	\begin{tabular}{cccccr}
	\hline\hline
 	 $J$ & $K$ & Sym. &  Experiment & Calculated & Obs$-$calc \\ 
	\hline	
	0 & 0 & $A_1$ &   0.00000 &   0.00000 &  0.00000 \\[-1mm]
	1 & 1 & $F_1$ &   5.71801 &   5.71800 &  0.00001 \\[-1mm]
	2 & 2 & $E$   &  17.15306 &  17.15302 &  0.00004 \\[-1mm]
	2 & 1 & $F_2$ &  17.15321 &  17.15317 &  0.00004 \\[-1mm]
	3 & 2 & $A_2$ &  34.30453 &  34.30448 &  0.00005 \\[-1mm]
	3 & 3 & $F_1$ &  34.30319 &  34.30313 &  0.00006 \\[-1mm]
	3 & 1 & $F_2$ &  34.30379 &  34.30373 &  0.00006 \\[-1mm]
	4 & 0 & $A_1$ &  57.16474 &  57.16467 &  0.00007 \\[-1mm]
	4 & 2 & $E$   &  57.16653 &  57.16647 &  0.00006 \\[-1mm]
	4 & 1 & $F_1$ &  57.16578 &  57.16572 &  0.00006 \\[-1mm]
	4 & 3 & $F_2$ &  57.16877 &  57.16872 &  0.00005 \\[-1mm]
	5 & 2 & $E$   &  85.74233 &  85.74231 &  0.00002 \\[-1mm]
	5 & 1 & $F_1$ &  85.73510 &  85.73504 &  0.00006 \\[-1mm]
	5 & 3 & $F_1$ &  85.74330 &  85.74328 &  0.00002 \\[-1mm]
	5 & 5 & $F_2$ &  85.73711 &  85.73707 &  0.00004 \\[-1mm]
	6 & 4 & $A_1$ & 120.02574 & 120.02581 & -0.00007 \\[-1mm]
	6 & 2 & $A_2$ & 120.01143 & 120.01144 & -0.00001 \\[-1mm]
	6 & 6 & $E$   & 120.00784 & 120.00784 &  0.00000 \\[-1mm]
    6 & 3 & $F_1$ & 120.02350 & 120.02356 & -0.00006 \\[-1mm]
    6 & 1 & $F_2$ & 120.00873 & 120.00874 & -0.00001 \\[-1mm]
    6 & 5 & $F_2$ & 120.02097 & 120.02102 & -0.00005 \\
    \hline\hline
    \end{tabular}
\end{center}
\end{table}

\subsection{Vibrational $J=0$ energies}
\label{sec:J0_energies}

 To calculate rovibrational energy levels, transition frequencies and corresponding intensities we use the variational nuclear motion code TROVE.~\cite{TROVE2007} Here we only summarize the key aspects of our calculations. Details of the general methodology can be found in Refs.~\onlinecite{TROVE2007,YuBaYa09.NH3,15YaYu.ADF}.

 The rovibrational Hamiltonian was represented as a power series expansion around the equilibrium geometry in terms of the coordinates given in Eqs.~\eqref{eq:stretch} to \eqref{eq:ang5}, and was constructed numerically using an automatic differentiation method.~\cite{15YaYu.ADF} The kinetic and potential energy operators were truncated at 6th and 8th order, respectively, which is sufficient for our purposes. For a discussion of the associated errors of such a scheme see Refs.~\onlinecite{TROVE2007,15YaYu.ADF}. Note that atomic mass values were employed in the subsequent TROVE calculations.
 
 The vibrational basis set was generated using a multi-step contraction scheme. For SiH$_{4}$ the polyad number
\begin{equation}\label{eq:polyad}
P = 2(n_1+n_2+n_3+n_4)+n_5+n_6+n_7+n_8+n_9 \leq P_{\mathrm{max}}
\end{equation}

\noindent controls the size of the basis set and does not exceed a predefined maximum value $P_{\mathrm{max}}$. For $J=0$ vibrational energy level calculations we set $P_{\mathrm{max}}=14$. Here the quantum numbers $n_k$ for $k=1,\ldots,9$ correspond to primitive basis functions $\phi_{n_k}$, which are obtained by solving a one-dimensional Schr\"{o}dinger equation for each vibrational mode by means of the Numerov-Cooley method.~\cite{Numerov1924,Cooley1961}

 The normal modes of silane are classified by the symmetry species, $A_{1}$, $E$, and $F_2$. Of $A_{1}$ symmetry is the non-degenerate symmetric stretching mode $\nu_{1}$ ($2186.87{\,}$cm$^{-1}$). The doubly degenerate asymmetric bending mode $\nu_{2}$ ($970.93{\,}$cm$^{-1}$) has $E$ symmetry. Whilst of $F_2$ symmetry are the triply degenerate modes; the asymmetric stretching mode $\nu_{3}$ ($2189.19{\,}$cm$^{-1}$), and the asymmetric bending mode $\nu_{4}$ ($913.47{\,}$cm$^{-1}$). The values in parentheses are the experimentally determined values from Ref.~\onlinecite{STDS:1998}. To be of spectroscopic use we map the vibrational quantum numbers $n_k$ of TROVE to the normal mode quantum numbers $\mathrm{v}_k$ commonly used. For SiH$_4$ the vibrational states are labelled as $\mathrm{v_1}\nu_1+\mathrm{v_2}\nu_2+\mathrm{v_3}\nu_3+\mathrm{v_4}\nu_4$ where $\mathrm{v}_i$ counts the level of excitation.

 In Table~\ref{tab:j0_28sih4} the computed vibrational energies using the CBS-F12$^{\,\mathrm{HL}}$ PES are listed against all available experimental data up to $8500\,$cm$^{-1}$. The four fundamental frequencies are all reproduced with sub-wavenumber accuracy, resulting in an overall rms error of $0.63\,$cm$^{-1}$ and a mean-absolute-deviation (mad) of $0.57\,$cm$^{-1}$. Altogether the 49 experimental levels are reproduced with a rms error of $1.33\,$cm$^{-1}$ and mad of $1.07\,$cm$^{-1}$. Note that energies are converged to $0.01\,$cm$^{-1}$ or better (the majority are converged to orders of magnitude lower), except for the two levels at $8347.86\,$cm$^{-1}$ which are converged to within $0.02\,$cm$^{-1}$. This was confirmed by performing a complete vibrational basis set extrapolation with values of $P_{\mathrm{max}}=\lbrace 10,12,14\rbrace$ (see Refs.~\onlinecite{OvThYu08.PH3,15OwYuYa.CH3Cl} for further details).

\setlength\LTleft{0pt}
\setlength\LTright{0pt}
\LTcapwidth=\textwidth
\begin{longtable*}[ht]{@{\extracolsep{0.4cm}} l c c c c c}
\caption{\label{tab:j0_28sih4}Comparison of calculated and experimental $J=0$ vibrational term values (in cm$^{-1}$) for $^{28}$SiH$_4$. The zero-point energy was computed to be $6847.084\,$cm$^{-1}$.}\\ \hline\hline
Mode & Sym. & Experiment & Calculated & Obs$-$calc & Ref.\\ \hline
\endfirsthead
\caption{(\textit{Continued})}\\ \hline 
Mode & Sym. & Experiment & Calculated & Obs$-$calc & Ref.\\ \hline
\endhead
$\nu_4$       & $F_2$ &  913.47 &  912.85 &  0.62 & \onlinecite{STDS:1998}\\[-1mm]
$\nu_2$       & $E  $ &  970.93 &  970.14 &  0.79 & \onlinecite{STDS:1998}\\[-1mm]
$2\nu_4$      & $A_1$ & 1811.80 & 1810.90 &  0.90 & \onlinecite{STDS:1998}\\[-1mm]
$2\nu_4$      & $F_2$ & 1824.19 & 1823.15 &  1.04 & \onlinecite{STDS:1998}\\[-1mm]
$2\nu_4$      & $E$   & 1827.81 & 1827.00 &  0.81 & \onlinecite{STDS:1998}\\[-1mm]
$\nu_2+\nu_4$ & $F_2$ & 1881.96 & 1880.87 &  1.09 & \onlinecite{STDS:1998}\\[-1mm]
$\nu_2+\nu_4$ & $F_1$ & 1887.10 & 1885.36 &  1.74 & \onlinecite{STDS:1998}\\[-1mm]
$2\nu_2$      & $A_1$ & 1937.50 & 1935.84 &  1.66 & \onlinecite{STDS:1998}\\[-1mm]
$2\nu_2$      & $E$   & 1942.77 & 1941.29 &  1.48 & \onlinecite{STDS:1998}\\[-1mm]
$\nu_1$       & $A_1$ & 2186.87 & 2187.63 & -0.76 & \onlinecite{STDS:1998}\\[-1mm]
$\nu_3$       & $F_2$ & 2189.19 & 2189.32 & -0.13 & \onlinecite{STDS:1998}\\[-1mm]
$3\nu_4$      & $F_2$ & 2713.07 & 2712.16 &  0.91 & \onlinecite{STDS:1998}\\[-1mm]
$3\nu_4$      & $A_1$ & 2731.17 & 2729.97 &  1.20 & \onlinecite{STDS:1998}\\[-1mm]
$3\nu_4$      & $F_1$ & 2735.42 & 2734.26 &  1.16 & \onlinecite{STDS:1998}\\[-1mm]
$3\nu_4$      & $F_2$ & 2739.35 & 2738.48 &  0.87 & \onlinecite{STDS:1998}\\[-1mm]
$\nu_2+2\nu_4$& $E$   & 2780.47 & 2779.32 &  1.15 & \onlinecite{STDS:1998}\\[-1mm]
$\nu_2+2\nu_4$& $F_1$ & 2793.32 & 2791.84 &  1.48 & \onlinecite{STDS:1998}\\[-1mm]
$\nu_2+2\nu_4$& $A_1$ & 2795.11 & 2793.94 &  1.17 & \onlinecite{STDS:1998}\\[-1mm]
$\nu_2+2\nu_4$& $F_2$ & 2797.41 & 2795.53 &  1.88 & \onlinecite{STDS:1998}\\[-1mm]
$\nu_2+2\nu_4$& $E$   & 2800.20 & 2798.25 &  1.95 & \onlinecite{STDS:1998}\\[-1mm]
$\nu_2+2\nu_4$& $A_2$ & 2803.95 & 2801.56 &  2.39 & \onlinecite{STDS:1998}\\[-1mm]
$2\nu_2+\nu_4$& $F_2$ & 2848.26 & 2846.60 &  1.66 & \onlinecite{STDS:1998}\\[-1mm]
$2\nu_2+\nu_4$& $F_1$ & 2856.43 & 2854.36 &  2.07 & \onlinecite{STDS:1998}\\[-1mm]
$2\nu_2+\nu_4$& $F_2$ & 2859.74 & 2857.18 &  2.56 & \onlinecite{STDS:1998}\\[-1mm]
$3\nu_2$      & $E$   & 2904.99 & 2902.60 &  2.39 & \onlinecite{STDS:1998}\\[-1mm]
$3\nu_2$      & $A_1$ & 2915.40 & 2913.34 &  2.06 & \onlinecite{STDS:1998}\\[-1mm]
$3\nu_2$      & $A_2$ & 2915.48 & 2913.44 &  2.04 & \onlinecite{STDS:1998}\\[-1mm]
$\nu_3+\nu_4$ & $F_1$ & 3094.81 & 3094.35 &  0.46 & \onlinecite{00WaSixx.SiH4}$^{\,a}$\\[-1mm]
$\nu_1+\nu_4$ & $F_2$ & 3095.26 & 3095.10 &  0.16 & \onlinecite{00WaSixx.SiH4}$^{\,a}$\\[-1mm]
$\nu_3+\nu_4$ & $E$   & 3095.86 & 3095.52 &  0.34 & \onlinecite{00WaSixx.SiH4}$^{\,a}$\\[-1mm]
$\nu_3+\nu_4$ & $F_2$ & 3098.02 & 3097.60 &  0.42 & \onlinecite{00WaSixx.SiH4}$^{\,a}$\\[-1mm]
$\nu_3+\nu_4$ & $A_1$ & 3099.48 & 3098.73 &  0.75 & \onlinecite{00WaSixx.SiH4}$^{\,a}$\\[-1mm]
$\nu_2+\nu_3$ & $F_2$ & 3152.59 & 3152.92 & -0.33 & \onlinecite{00WaSixx.SiH4}$^{\,a}$\\[-1mm]
$\nu_2+\nu_3$ & $F_1$ & 3153.08 & 3152.17 &  0.91 & \onlinecite{00WaSixx.SiH4}$^{\,a}$\\[-1mm]
$\nu_1+\nu_2$ & $E$   & 3153.60 & 3152.12 &  1.48 & \onlinecite{00WaSixx.SiH4}$^{\,a}$\\[-1mm]
$2\nu_3$      & $A_1$ & 4308.87 & 4308.96 & -0.09 & \onlinecite{94ZhCaSt.SiH4}$^{\,b}$\\[-1mm]
$\nu_1+\nu_3$ & $F_2$ & 4309.35 & 4309.89 & -0.54 & \onlinecite{91ZhQiMa.SiH4}\\[-1mm]
$2\nu_1$      & $A_1$ & 4374.56 & 4375.92 & -1.36 & \onlinecite{97PeCaxx.SiH4}$^{\,c}$\\[-1mm]
$2\nu_3$      & $E$   & 4378.40 & 4380.23 & -1.83 & \onlinecite{STDS:1998}\\[-1mm]
$2\nu_3$      & $F_2$ & 4380.28 & 4378.73 &  1.55 & \onlinecite{97PeCaxx.SiH4}$^{\,c}$\\[-1mm]
$\nu_1+2\nu_3$& $A_1$ & 6362.05 & 6362.88 & -0.83 & \onlinecite{94ZhCaSt.SiH4}$^{\,d}$\\[-1mm]
$3\nu_3$      & $F_2$ & 6362.05 & 6362.97 & -0.92 & \onlinecite{94ZhCaSt.SiH4}$^{\,d}$\\[-1mm]
$3\nu_1$      & $A_1$ & 6496.13 & 6498.19 & -2.06 & \onlinecite{97PeCaxx.SiH4}$^{\,c}$\\[-1mm]
$2\nu_1+\nu_3$& $F_2$ & 6497.45 & 6498.48 & -1.03 & \onlinecite{95SuWaZh.SiH4}\\[-1mm]
$\nu_1+2\nu_3$& $E$   & 6500.30 & 6500.58 & -0.28 & \onlinecite{97PeCaxx.SiH4}$^{\,c}$\\[-1mm]
$3\nu_3$      & $F_2$ & 6500.60 & 6500.71 & -0.11 & \onlinecite{97PeCaxx.SiH4}$^{\,c}$\\[-1mm]
$3\nu_3$      & $F_1$ & 6502.88 & 6502.94 & -0.06 & \onlinecite{97PeCaxx.SiH4}$^{\,c}$\\[-1mm]
$\nu_1+3\nu_3$& $A_1$ & 8347.86 & 8349.38 & -1.52 & \onlinecite{94ZhCaSt.SiH4}$^{\,d}$\\[-1mm]
$\nu_1+3\nu_3$& $F_2$ & 8347.86 & 8349.39 & -1.53 & \onlinecite{94ZhCaSt.SiH4}$^{\,d}$\\
\hline\hline
\caption*{
$^a$ Originally attributed to Ref.~\onlinecite{STDS:1998}, but unable to confirm value independently.\\[-2mm]
$^b$ Originally attributed to Ref.~\onlinecite{91ZhQiMa.SiH4}. $^c$ Originally attributed to Ref.~\onlinecite{Chevalier:1988}.\\[-2mm]
$^d$ Originally attributed to Refs.~\onlinecite{89ZhZhMa.SiH4,90ZhZhMa.SiH4,90ZhMaZh.SiH4}.\\
}
\end{longtable*}

 Of the 35 term values up to $3153.60{\,}$cm$^{-1}$, the energy of 32 levels is underestimated by the CBS-F12$^{\,\mathrm{HL}}$ PES. This can be explained by the residual errors of the $\nu_2$ and $\nu_4$ fundamentals, which largely dictates the accuracy of the subsequent combination bands and overtones. Above $3153.60{\,}$cm$^{-1}$ computed energy levels are consistently higher than experiment which is a result of overestimating the $\nu_1$ and $\nu_3$ fundamentals. Despite this, the performance of the CBS-F12$^{\,\mathrm{HL}}$ PES is extremely encouraging, especially considering that for vibrational $J=0$ energy levels the PES can be regarded as an \textit{ab initio} surface.
 
 Experimental values for stretching overtones above $8500{\,}$cm$^{-1}$ are available.~\cite{84BeLaOk.SiH4,90ZhZhMa.SiH4,94ZhCaSt.SiH4} However, the corresponding values in TROVE are harder to identify given the increased density of states at higher energies. Highly excited modes also show slower convergence with respect to vibrational basis set size. Thus, to obtain reasonably well converged energies would require calculations with $P_{\mathrm{max}}=16$ or greater, which is currently unachievable with the computational resources available to us.
 
 As an aside in Table~\ref{tab:j0_abeq} we show the effect of the empirical refinement of the equilibrium geometry on the fundamental frequencies. Results computed using the \textit{ab initio} bond length (overall rms error of $0.57\,$cm$^{-1}$) are marginally better which is to be expected. In the refined geometry PES the shape of the original \textit{ab initio} PES has been altered by shifting its minimum, resulting in a poorer representation of vibrational energies. For spectral analysis an improved description of rotational structure is more desirable however, as vibrational band position can be easily corrected at a later stage.~\cite{YuBaYa09.NH3}
 
\begin{table}[h!]
\tabcolsep=5pt
\caption{\label{tab:j0_abeq}Comparison of the computed fundamental term values (in cm$^{-1}$) with the refined and \textit{ab initio} equilibrium geometry.}
\begin{center}
	\begin{tabular}{l c c c c c c}
	\hline\hline
 	 Mode & Sym. & Experiment$^a$ & Refined eq. (A) & \textit{Ab initio} eq. (B) & Obs-calc (A) & Obs-calc (B) \\ 
	\hline	
     $\nu_1$ & $A_1$ & 2186.87 & 2187.63 & 2187.63 & -0.76 & -0.76\\[-1mm]
     $\nu_2$ & $E  $ &  970.93 &  970.14 &  970.26 &  0.79 &  0.67\\[-1mm]
     $\nu_3$ & $F_2$ & 2189.19 & 2189.32 & 2189.31 & -0.13 & -0.12\\[-1mm]
     $\nu_4$ & $F_2$ &  913.47 &  912.85 &  912.97 &  0.62 &  0.50\\
    \hline\hline
    \end{tabular}
\end{center}
$^a$ See Table~\ref{tab:j0_28sih4} for experimental references.\\
\end{table}

\subsection{Vibrational transition moments}
\label{sec:vib_tm}

 The vibrational transition moment is defined as,
\begin{equation}
\mu_{if} = \sqrt{\sum_{\alpha=x,y,z}{\lvert\bra{\Phi^{(f)}_{\mathrm{vib}}}\bar{\mu}_{\alpha}\ket{\Phi^{(i)}_{\mathrm{vib}}}\rvert}^2}
\end{equation}

\noindent where $\ket{\Phi^{(i)}_{\mathrm{vib}}}$ and $\ket{\Phi^{(f)}_{\mathrm{vib}}}$ are the initial and final state vibrational eigenfunctions respectively, and $\bar{\mu}_{\alpha}$ is the electronically averaged dipole moment function along the molecule-fixed axis $\alpha=x,y,z$. In Table~\ref{tab:TM} we list computed vibrational transition moments from the vibrational ground state. Calculations used the CBS-F12$^{\,\mathrm{HL}}$ PES and a polyad number of $P_{\mathrm{max}}=12$ which ensured converged results.

\begin{table}[h!]
\tabcolsep=0.25cm
\caption{\label{tab:TM}Calculated vibrational transition moments (in Debye) and frequencies (in cm$^{-1}$) from the vibrational ground state for $^{28}$SiH$_4$. Only levels of $F_2$ symmetry are accessible from the ground state in IR absorption.}
\begin{center}
	\begin{tabular}{l c c c c}
	\hline\hline
 	 Mode & Sym. & Experiment$^a$ & Calculated & $\mu_{if}$\\ 
	\hline	
     $\nu_4$        & $F_2$ &  913.47 &  912.85 & 0.4149E+0 \\[-1mm]
     $2\nu_4$       & $F_2$ & 1824.19 & 1823.15 & 0.2500E-2 \\[-1mm]
     $\nu_2+\nu_4$  & $F_2$ & 1881.96 & 1880.87 & 0.2350E-1 \\[-1mm]
     $\nu_3$        & $F_2$ & 2189.19 & 2189.32 & 0.2470E+0\\[-1mm]
     $3\nu_4$       & $F_2$ & 2713.07 & 2712.16 & 0.4578E-2 \\[-1mm]
     $3\nu_4$       & $F_2$ & 2739.35 & 2738.48 & 0.8123E-3 \\[-1mm]
     $\nu_2+2\nu_4$ & $F_2$ & 2797.41 & 2795.53 & 0.1734E-2 \\[-1mm]
     $2\nu_2+\nu_4$ & $F_2$ & 2848.26 & 2846.60 & 0.1835E-2 \\[-1mm]
     $2\nu_2+\nu_4$ & $F_2$ & 2859.74 & 2857.18 & 0.9093E-4 \\[-1mm]
     $\nu_1+\nu_4$  & $F_2$ & 3095.26 & 3095.10 & 0.1320E-1 \\[-1mm]
     $\nu_3+\nu_4$  & $F_2$ & 3098.02 & 3097.60 & 0.1319E-1 \\[-1mm]
     $\nu_2+\nu_3$  & $F_2$ & 3152.59 & 3152.92 & 0.1050E-1 \\[-1mm]
     $4\nu_4$       & $F_2$ &   -     & 3609.08 & 0.4741E-3 \\[-1mm]
     $4\nu_4$       & $F_2$ &   -     & 3638.92 & 0.1892E-4 \\[-1mm]
     $\nu_2+3\nu_4$ & $F_2$ &   -     & 3677.72 & 0.6075E-3 \\[-1mm]
     $\nu_2+3\nu_4$ & $F_2$ &   -     & 3704.01 & 0.5424E-3 \\[-1mm]
     $\nu_2+3\nu_4$ & $F_2$ &   -     & 3707.66 & 0.2098E-4 \\[-1mm]
     $2\nu_2+2\nu_4$& $F_2$ &   -     & 3758.50 & 0.1628E-3 \\[-1mm]
     $2\nu_2+2\nu_4$& $F_2$ &   -     & 3767.13 & 0.5799E-4 \\[-1mm]
     $3\nu_2+\nu_4$ & $F_2$ &   -     & 3810.86 & 0.2432E-3 \\[-1mm]
     $3\nu_2+\nu_4$ & $F_2$ &   -     & 3827.61 & 0.3848E-3 \\[-1mm]
     $\nu_1+\nu_3$  & $F_2$ & 4309.35 & 4309.89 & 0.1336E-1 \\[-1mm]
     $2\nu_3$       & $F_2$ & 4380.28 & 4378.73 & 0.4262E-2 \\[-1mm]
     $3\nu_3$       & $F_2$ & 6362.05 & 6362.97 & 0.5762E-3 \\[-1mm]
     $2\nu_1+\nu_3$ & $F_2$ & 6497.45 & 6498.48 & 0.5813E-3 \\[-1mm]
     $3\nu_3$       & $F_2$ & 6500.60 & 6500.71 & 0.1517E-3 \\[-1mm]
     $\nu_1+3\nu_3$ & $F_2$ & 8347.86 & 8349.39 & 0.1390E-2 \\
    \hline\hline
    \end{tabular}
\end{center}
$^a$ See Table~\ref{tab:j0_28sih4} for experimental references.\\
\end{table}

 Experimentally determined transitions moments have only been derived for the $\nu_3$ ($2189.19{\,}$cm$^{-1}$) and $\nu_4$ ($913.47{\,}$cm$^{-1}$) modes. \citet{76FoPexx.SiH4} using earlier band intensity measurements~\cite{62BaMcxx.SiH4,62LeKixx.SiH4} found $\mu_{\nu_3}=0.139\pm 4\%{\,}$D and $\mu_{\nu_4}=0.232\pm 7\%{\,}$D. The reliability of the intensity data~\cite{62BaMcxx.SiH4,62LeKixx.SiH4} has however been questioned.~\cite{94CoMcSt.SiH4} In other work, \citet{92Caxxxx.SiH4} determined a transition moment of $\mu_{\nu_3}=0.1293\pm 3\%{\,}$D. Whilst a value of $\mu_{\nu_4}=0.247{\,}$D was quoted in Ref.~\onlinecite{93KeRixx.SiH4} but attributed to unpublished results.

 Although the experimental situation is not entirely clear, the computed TROVE transition moments of $\mu_{\nu_3}=0.2470{\,}$D and $\mu_{\nu_4}=0.4149{\,}$D are notably larger than their experimental counterparts. We will show in Sec.~\ref{sec:nu_3} and Sec.~\ref{sec:overview} that our DMS does marginally overestimate the strength of line intensities. The magnitude of this overestimation is not consistent with the discrepancy in the experimental and computed values for $\mu_{\nu_3}$ and $\mu_{\nu_4}$ however. Experimentally derived transition moments for the other levels of silane could help clarify previous results and assist future theoretical benchmarking. 
 
 It is worth nothing that if we use the values from Ref.~\onlinecite{76FoPexx.SiH4} and compare the ratio $\mu_{\nu_3}^{\mathrm{exp}}/\mu_{\nu_4}^{\mathrm{exp}}=0.599$ with $\mu_{\nu_3}^{\mathrm{TROVE}}/\mu_{\nu_4}^{\mathrm{TROVE}}=0.595$, there is excellent agreement which suggests our relative intensity for the two strongest bands is reasonable. 

\subsection{Absolute line intensities of the $\nu_3$ band}
\label{sec:nu_3}
 
 To simulate absolute absorption intensities we use the expression,
\begin{equation}
I(f \leftarrow i) = \frac{A_{if}}{8\pi c}g_{\mathrm{ns}}(2 J_{f}+1)\frac{\exp\left(-E_{i}/kT\right)}{Q(T)\; \nu_{if}^{2}}\left[1-\exp\left(-\frac{hc\nu_{if}}{kT}\right)\right] ,
\end{equation}

\noindent where $A_{if}$ is the Einstein-A coefficient of a transition with frequency $\nu_{if}$ between an initial state with energy $E_i$, and a final state with rotational quantum number $J_f$. Here $k$ is the Boltzmann constant, $T$ is the absolute temperature, and $c$ is the speed of light. The nuclear spin statistical weights are $g_{\mathrm{ns}}=\lbrace 5,5,2,3,3\rbrace$ for states of symmetry $\lbrace A_1,A_2,E,F_1,F_2\rbrace$, respectively. The partition function $Q(T)$ was estimated using, $Q(T)\approx Q_{\mathrm{rot}}(T)\times Q_{\mathrm{vib}}(T)$. For tetrahedral molecules the rotational partition function is given as,~\cite{Fox:1970}
\begin{equation}
Q_{\mathrm{rot}}(T)=\frac{4}{3}\pi^{1/2}\left(\frac{Bhc}{kT}\right)^{-3/2}\exp\left(\frac{Bhc}{4kT}\right)
\end{equation}

\noindent where for SiH$_4$ we use a ground state rotational constant of $B=2.859$, which is consistent with Refs.~\onlinecite{74DaPiSa.SiH4,75PiGuAm.SiH4,84PiVaHe.SiH4}. At $T=296\,$K, $Q_{\mathrm{rot}}=1447.6001$, the vibrational partition function $Q_{\mathrm{vib}}=1.0551$,~\cite{STDS:1998} resulting in $Q=1527.3629$.

\begin{figure*}
\includegraphics[width=\textwidth,angle=0]{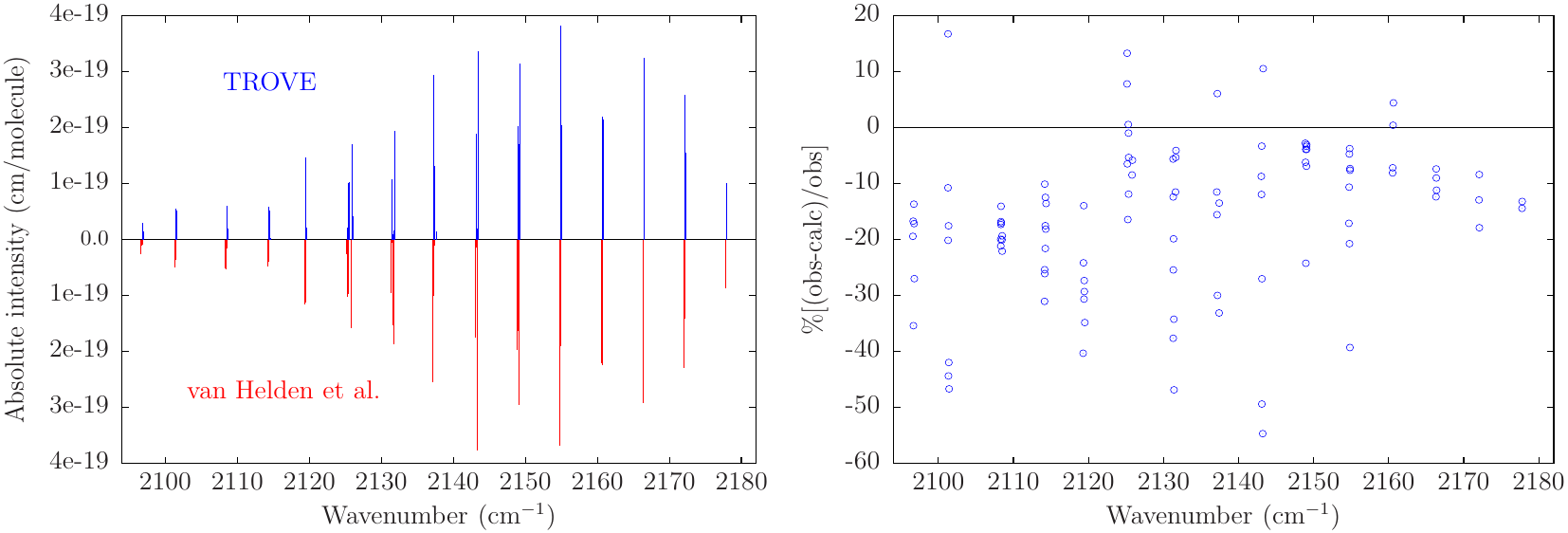}
\caption{\label{fig:v3_band}Absolute line intensities of the $\nu_3$ band for transitions up to $J=16$ (left) and the corresponding residuals $\left(\%\left[\frac{\mathrm{obs-calc}}{\mathrm{obs}}\right]\right)$ (right) when compared with measurements from \citet{15HeLoNa.SiH4}.}
\end{figure*}

 A recent high-resolution study of the $\nu_3$ band measured the absolute line intensities of numerous P-branch transitions up to $J=16$ at $296{\,}$K.~\cite{15HeLoNa.SiH4} Line intensities were recorded at a resolution of $0.0011{\,}$cm$^{-1}$ and were given an estimated experimental measurement accuracy of $10\%$. To validate our DMS and to a lesser extent the PES, in Table~\ref{tab:v3_intens} we compare frequencies and absolute line intensities of over 100 transitions from Ref.~\onlinecite{15HeLoNa.SiH4}. The results are also illustrated in Fig.~\eqref{fig:v3_band}.
  
\setlength\LTleft{0pt}
\setlength\LTright{0pt}
\LTcapwidth=\textwidth
\begin{longtable*}[ht]{@{\extracolsep{0.25cm}} c c c c c c c c c c c c}
\caption{\label{tab:v3_intens}Comparison of calculated and observed frequencies (in cm$^{-1}$) and absolute line intensities (in cm/molecule) for transitions between the $\nu_3$ and ground vibrational state. To quantify the error in the computed line intensity we use the percentage measure, $\%[\mathrm{(obs-calc)/obs}]$.}\\ \hline\hline
$\Gamma\p$ & $J\p$ & $K\p$ & $\Gamma\pp$ & $J\pp$ & $K\pp$ & $\nu_{\mathrm{obs}}$ & $\nu_{\mathrm{calc}}$ & $\Delta_{\mathrm{obs - calc}}$ & $I_{\mathrm{obs}}$ & $I_{\mathrm{calc}}$ & $\%\left[\frac{\mathrm{obs - calc}}{\mathrm{obs}}\right]$ \\ \hline
\endfirsthead
\caption{(\textit{Continued})}\\ \hline 
$\Gamma\p$ & $J\p$ & $K\p$ & $\Gamma\pp$ & $J\pp$ & $K\pp$ & $\nu_{\mathrm{obs}}$ & $\nu_{\mathrm{calc}}$ & $\Delta_{\mathrm{obs - calc}}$ & $S_{\mathrm{obs}}$ & $S_{\mathrm{calc}}$ & $\%\left[\frac{\mathrm{obs - calc}}{\mathrm{obs}}\right]$ \\ \hline
\endhead
$F_1$ & 1 & 1 & $F_2$ & 2 & 1 & 2177.782 & 2177.908 & -0.126 & 8.784E-20 & 1.005E-19 & -14.42 \\[-1.5mm]
$E$   & 1 & 1 & $E$   & 2 & 1 & 2177.793 & 2177.921 & -0.128 & 5.920E-20 & 6.701E-20 & -13.19 \\[-1.5mm]
$A_1$ & 2 & 1 & $A_2$ & 3 & 1 & 2172.045 & 2172.170 & -0.125 & 2.290E-19 & 2.586E-19 & -12.92 \\[-1.5mm]
$F_1$ & 2 & 1 & $F_2$ & 3 & 1 & 2172.072 & 2172.197 & -0.125 & 1.417E-19 & 1.535E-19 & -8.38 \\[-1.5mm]
$F_2$ & 2 & 1 & $F_1$ & 3 & 2 & 2172.091 & 2172.216 & -0.125 & 1.315E-19 & 1.550E-19 & -17.88 \\[-1.5mm]
$F_1$ & 3 & 1 & $F_2$ & 4 & 2 & 2166.306 & 2166.431 & -0.125 & 1.682E-19 & 1.889E-19 & -12.35 \\[-1.5mm]
$E$   & 3 & 1 & $E$   & 4 & 1 & 2166.340 & 2166.466 & -0.126 & 1.212E-19 & 1.301E-19 & -7.37 \\[-1.5mm]
$F_2$ & 3 & 2 & $F_1$ & 4 & 1 & 2166.357 & 2166.483 & -0.126 & 1.776E-19 & 1.935E-19 & -8.98 \\[-1.5mm]
$A_2$ & 3 & 2 & $A_1$ & 4 & 0 & 2166.377 & 2166.504 & -0.127 & 2.923E-19 & 3.250E-19 & -11.17 \\[-1.5mm]
$F_2$ & 4 & 2 & $F_1$ & 5 & 2 & 2160.524 & 2160.654 & -0.130 & 1.959E-19 & 2.118E-19 & -8.09 \\[-1.5mm]
$E$   & 4 & 2 & $E$   & 5 & 1 & 2160.547 & 2160.678 & -0.131 & 1.270E-19 & 1.361E-19 & -7.16 \\[-1.5mm]
$F_1$ & 4 & 1 & $F_2$ & 5 & 3 & 2160.591 & 2160.718 & -0.127 & 2.201E-19 & 2.191E-19 &  0.44 \\[-1.5mm]
$F_2$ & 4 & 1 & $F_1$ & 5 & 1 & 2160.629 & 2160.755 & -0.126 & 2.250E-19 & 2.150E-19 &  4.43 \\[-1.5mm]
$A_2$ & 5 & 2 & $A_1$ & 6 & 2 & 2154.706 & 2154.832 & -0.126 & 3.217E-19 & 3.766E-19 & -17.08 \\[-1.5mm]
$F_2$ & 5 & 2 & $F_1$ & 6 & 2 & 2154.738 & 2154.865 & -0.127 & 1.930E-19 & 2.135E-19 & -10.62 \\[-1.5mm]
$F_1$ & 5 & 1 & $F_2$ & 6 & 3 & 2154.768 & 2154.895 & -0.127 & 1.852E-19 & 1.939E-19 & -4.70 \\[-1.5mm]
$F_1$ & 5 & 1 & $F_2$ & 6 & 1 & 2154.780 & 2154.907 & -0.127 & 1.705E-20 & 2.058E-20 & -20.73 \\[-1.5mm]
$A_1$ & 5 & 3 & $A_2$ & 6 & 1 & 2154.810 & 2154.935 & -0.125 & 3.688E-19 & 3.826E-19 & -3.73 \\[-1.5mm]
$F_1$ & 5 & 1 & $F_2$ & 6 & 3 & 2154.844 & 2154.975 & -0.131 & 1.071E-20 & 1.491E-20 & -39.28 \\[-1.5mm]
$F_1$ & 5 & 1 & $F_2$ & 6 & 1 & 2154.856 & 2154.987 & -0.131 & 1.901E-19 & 2.039E-19 & -7.29 \\[-1.5mm]
$E$   & 5 & 1 & $E$   & 6 & 3 & 2154.862 & 2154.992 & -0.130 & 1.380E-19 & 1.485E-19 & -7.60 \\[-1.5mm]
$F_2$ & 6 & 2 & $F_1$ & 7 & 3 & 2148.893 & 2149.021 & -0.128 & 1.970E-19 & 2.025E-19 & -2.77 \\[-1.5mm]
$E$   & 6 & 2 & $E$   & 7 & 3 & 2148.926 & 2149.052 & -0.126 & 1.319E-19 & 1.400E-19 & -6.17 \\[-1.5mm]
$F_1$ & 6 & 3 & $F_2$ & 7 & 2 & 2148.954 & 2149.080 & -0.126 & 1.639E-19 & 1.702E-19 & -3.89 \\[-1.5mm]
$F_1$ & 6 & 3 & $F_2$ & 7 & 1 & 2148.976 & 2149.102 & -0.126 & 3.234E-20 & 4.017E-20 & -24.23 \\[-1.5mm]
$A_1$ & 6 & 3 & $A_2$ & 7 & 1 & 2149.046 & 2149.184 & -0.138 & 2.950E-19 & 3.153E-19 & -6.89 \\[-1.5mm]
$F_1$ & 6 & 1 & $F_2$ & 7 & 2 & 2149.052 & 2149.186 & -0.134 & 2.740E-20 & 2.832E-20 & -3.34 \\[-1.5mm]
$F_1$ & 6 & 1 & $F_2$ & 7 & 1 & 2149.074 & 2149.207 & -0.133 & 1.714E-19 & 1.781E-19 & -3.88 \\[-1.5mm]
$F_2$ & 6 & 3 & $F_1$ & 7 & 1 & 2149.082 & 2149.214 & -0.132 & 2.077E-19 & 2.140E-19 & -3.02 \\[-1.5mm]
$F_1$ & 7 & 3 & $F_2$ & 8 & 2 & 2143.025 & 2143.165 & -0.140 & 1.747E-19 & 1.899E-19 & -8.70 \\[-1.5mm]
$E$   & 7 & 3 & $E$   & 8 & 1 & 2143.056 & 2143.197 & -0.141 & 9.223E-20 & 1.032E-19 & -11.94 \\[-1.5mm]
$F_2$ & 7 & 2 & $F_1$ & 8 & 2 & 2143.084 & 2143.223 & -0.139 & 1.740E-19 & 1.798E-19 & -3.30 \\[-1.5mm]
$E$   & 7 & 3 & $E$   & 8 & 3 & 2143.104 & 2143.246 & -0.142 & 1.340E-20 & 2.002E-20 & -49.38 \\[-1.5mm]
$F_2$ & 7 & 2 & $F_1$ & 8 & 1 & 2143.125 & 2143.264 & -0.139 & 1.201E-20 & 1.525E-20 & -26.99 \\[-1.5mm]
$E$   & 7 & 1 & $E$   & 8 & 1 & 2143.228 & 2143.372 & -0.144 & 6.575E-21 & 1.017E-20 & -54.68 \\[-1.5mm]
$A_2$ & 7 & 1 & $A_1$ & 8 & 0 & 2143.286 & 2143.424 & -0.138 & 3.771E-19 & 3.373E-19 &  10.56 \\[-1.5mm]
$F_1$ & 8 & 3 & $F_2$ & 9 & 1 & 2137.100 & 2137.240 & -0.140 & 1.135E-20 & 1.265E-20 & -11.48 \\[-1.5mm]
$A_1$ & 8 & 3 & $A_2$ & 9 & 3 & 2137.136 & 2137.267 & -0.131 & 2.554E-19 & 2.951E-19 & -15.55 \\[-1.5mm]
$F_1$ & 8 & 2 & $F_2$ & 9 & 3 & 2137.173 & 2137.301 & -0.128 & 1.546E-19 & 1.452E-19 &  6.09 \\[-1.5mm]
$F_2$ & 8 & 2 & $F_1$ & 9 & 4 & 2137.198 & 2137.324 & -0.126 & 1.010E-19 & 1.313E-19 & -29.97 \\[-1.5mm]
$F_2$ & 8 & 2 & $F_1$ & 9 & 4 & 2137.417 & 2137.570 & -0.153 & 1.122E-20 & 1.493E-20 & -33.10 \\[-1.5mm]
$F_2$ & 8 & 1 & $F_1$ & 9 & 2 & 2137.426 & 2137.569 & -0.143 & 3.185E-21 & 3.613E-21 & -13.46 \\[-1.5mm]
$E$   & 9 & 3 & $E$   & 10 & 3 & 2131.274 & 2131.402 & -0.128 & 8.116E-20 & 8.571E-20 & -5.61 \\[-1.5mm]
$F_2$ & 9 & 4 & $F_1$ & 10 & 3 & 2131.298 & 2131.424 & -0.126 & 9.629E-20 & 1.082E-19 & -12.38 \\[-1.5mm]
$A_1$ & 9 & 1 & $A_2$ & 10 & 1 & 2131.302 & 2131.439 & -0.137 & 4.663E-20 & 6.417E-20 & -37.62 \\[-1.5mm]
$F_1$ & 9 & 3 & $F_2$ & 10 & 2 & 2131.315 & 2131.445 & -0.130 & 5.899E-21 & 7.399E-21 & -25.42 \\[-1.5mm]
$F_2$ & 9 & 4 & $F_1$ & 10 & 1 & 2131.340 & 2131.467 & -0.127 & 2.822E-20 & 3.382E-20 & -19.86 \\[-1.5mm]
$F_1$ & 9 & 3 & $F_2$ & 10 & 1 & 2131.381 & 2131.512 & -0.131 & 1.117E-20 & 1.499E-20 & -34.22 \\[-1.5mm]
$E$   & 9 & 3 & $E$   & 10 & 5 & 2131.399 & 2131.527 & -0.128 & 6.337E-21 & 9.306E-21 & -46.85 \\[-1.5mm]
$F_1$ & 9 & 1 & $F_2$ & 10 & 4 & 2131.594 & 2131.678 & -0.084 & 6.694E-21 & 3.753E-21 &  43.93 \\[-1.5mm]
$F_2$ & 9 & 3 & $F_1$ & 10 & 3 & 2131.600 & 2131.764 & -0.164 & 1.449E-20 & 1.615E-20 & -11.48 \\[-1.5mm]
$A_2$ & 9 & 4 & $A_1$ & 10 & 4 & 2131.629 & 2131.796 & -0.167 & 1.534E-19 & 1.616E-19 & -5.31 \\[-1.5mm]
$A_1$ & 9 & 3 & $A_2$ & 10 & 1 & 2131.672 & 2131.826 & -0.154 & 1.876E-19 & 1.952E-19 & -4.06 \\[-1.5mm]
$F_2$ & 10 & 4 & $F_1$ & 11 & 2 & 2125.142 & 2125.281 & -0.139 & 1.315E-20 & 1.212E-20 &  7.82 \\[-1.5mm]
$E$   & 10 & 1 & $E$   & 11 & 3 & 2125.162 & 2125.302 & -0.140 & 2.551E-20 & 2.212E-20 & 13.30 \\[-1.5mm]
$F_2$ & 10 & 4 & $F_1$ & 11 & 4 & 2125.194 & 2125.333 & -0.139 & 1.512E-20 & 1.610E-20 & -6.46 \\[-1.5mm]
$E$   & 10 & 1 & $E$   & 11 & 1 & 2125.249 & 2125.389 & -0.140 & 8.867E-21 & 1.032E-20 & -16.41 \\[-1.5mm]
$F_2$ & 10 & 4 & $F_1$ & 11 & 2 & 2125.312 & 2125.441 & -0.129 & 1.016E-19 & 1.011E-19 &  0.58 \\[-1.5mm]
$E$   & 10 & 2 & $E$   & 11 & 3 & 2125.340 & 2125.467 & -0.127 & 5.186E-20 & 5.236E-20 & -0.97 \\[-1.5mm]
$F_1$ & 10 & 1 & $F_2$ & 11 & 2 & 2125.348 & 2125.481 & -0.133 & 1.369E-20 & 1.531E-20 & -11.88 \\[-1.5mm]
$F_1$ & 10 & 3 & $F_2$ & 11 & 3 & 2125.362 & 2125.488 & -0.126 & 9.684E-20 & 1.020E-19 & -5.32 \\[-1.5mm]
$A_1$ & 10 & 4 & $A_2$ & 11 & 1 & 2125.809 & 2125.973 & -0.164 & 1.579E-19 & 1.712E-19 & -8.44 \\[-1.5mm]
$E$   & 10 & 4 & $E$   & 11 & 1 & 2125.851 & 2126.025 & -0.174 & 3.963E-20 & 4.194E-20 & -5.82 \\[-1.5mm]
$F_2$ & 11 & 1 & $F_1$ & 12 & 4 & 2119.300 & 2119.431 & -0.131 & 9.978E-21 & 1.400E-20 & -40.30 \\[-1.5mm]
$A_2$ & 11 & 2 & $A_1$ & 12 & 4 & 2119.331 & 2119.461 & -0.130 & 1.160E-19 & 1.440E-19 & -24.15 \\[-1.5mm]
$F_1$ & 11 & 3 & $F_2$ & 12 & 5 & 2119.389 & 2119.515 & -0.126 & 6.041E-20 & 6.883E-20 & -13.94 \\[-1.5mm]
$A_1$ & 11 & 3 & $A_2$ & 12 & 3 & 2119.414 & 2119.540 & -0.126 & 1.131E-19 & 1.477E-19 & -30.63 \\[-1.5mm]
$F_2$ & 11 & 1 & $F_1$ & 12 & 2 & 2119.440 & 2119.571 & -0.131 & 1.284E-20 & 1.634E-20 & -27.32 \\[-1.5mm]
$F_1$ & 11 & 3 & $F_2$ & 12 & 1 & 2119.449 & 2119.576 & -0.127 & 7.866E-21 & 1.017E-20 & -29.29 \\[-1.5mm]
$F_2$ & 11 & 2 & $F_1$ & 12 & 2 & 2119.508 & 2119.635 & -0.127 & 1.635E-20 & 2.204E-20 & -34.82 \\[-1.5mm]
$F_2$ & 12 & 1 & $F_1$ & 13 & 1 & 2114.154 & 2114.321 & -0.167 & 4.479E-20 & 5.868E-20 & -31.03 \\[-1.5mm]
$E$   & 12 & 5 & $E$   & 13 & 1 & 2114.169 & 2114.352 & -0.183 & 2.707E-20 & 3.394E-20 & -25.38 \\[-1.5mm]
$F_1$ & 12 & 1 & $F_2$ & 13 & 1 & 2114.179 & 2114.349 & -0.170 & 4.882E-20 & 5.374E-20 & -10.08 \\[-1.5mm]
$F_2$ & 12 & 1 & $F_1$ & 13 & 2 & 2114.187 & 2114.373 & -0.186 & 3.173E-20 & 4.001E-20 & -26.09 \\[-1.5mm]
$A_1$ & 12 & 1 & $A_2$ & 13 & 5 & 2114.252 & 2114.453 & -0.201 & 4.283E-20 & 5.208E-20 & -21.59 \\[-1.5mm]
$F_2$ & 12 & 3 & $F_1$ & 13 & 5 & 2114.259 & 2114.457 & -0.198 & 2.253E-20 & 2.647E-20 & -17.51 \\[-1.5mm]
$F_1$ & 12 & 4 & $F_2$ & 13 & 1 & 2114.263 & 2114.463 & -0.200 & 2.538E-20 & 2.854E-20 & -12.45 \\[-1.5mm]
$A_2$ & 12 & 4 & $A_1$ & 13 & 2 & 2114.309 & 2114.506 & -0.197 & 3.990E-20 & 4.713E-20 & -18.13 \\[-1.5mm]
$F_2$ & 12 & 3 & $F_1$ & 13 & 2 & 2114.354 & 2114.554 & -0.200 & 2.886E-21 & 3.277E-21 & -13.55 \\[-1.5mm]
$E$   & 13 & 2 & $E$   & 14 & 7 & 2108.308 & 2108.486 & -0.178 & 2.272E-20 & 2.725E-20 & -19.96 \\[-1.5mm]
$F_1$ & 13 & 1 & $F_2$ & 14 & 1 & 2108.321 & 2108.499 & -0.178 & 3.210E-20 & 3.888E-20 & -21.14 \\[-1.5mm]
$A_2$ & 13 & 5 & $A_1$ & 14 & 6 & 2108.343 & 2108.545 & -0.202 & 5.088E-20 & 5.941E-20 & -16.77 \\[-1.5mm]
$F_1$ & 13 & 2 & $F_2$ & 14 & 2 & 2108.349 & 2108.544 & -0.195 & 2.889E-20 & 3.389E-20 & -17.30 \\[-1.5mm]
$A_1$ & 13 & 2 & $A_2$ & 14 & 1 & 2108.354 & 2108.535 & -0.181 & 5.234E-20 & 5.969E-20 & -14.04 \\[-1.5mm]
$F_2$ & 13 & 2 & $F_1$ & 14 & 3 & 2108.392 & 2108.590 & -0.198 & 2.090E-20 & 2.445E-20 & -17.00 \\[-1.5mm]
$F_1$ & 13 & 2 & $F_2$ & 14 & 5 & 2108.482 & 2108.694 & -0.212 & 1.629E-20 & 1.955E-20 & -20.03 \\[-1.5mm]
$F_2$ & 13 & 3 & $F_1$ & 14 & 1 & 2108.501 & 2108.711 & -0.210 & 1.259E-20 & 1.537E-20 & -22.06 \\[-1.5mm]
$E$   & 13 & 4 & $E$   & 14 & 3 & 2108.510 & 2108.721 & -0.211 & 9.767E-21 & 1.165E-20 & -19.31 \\[-1.5mm]
$A_1$ & 14 & 3 & $A_2$ & 15 & 5 & 2101.289 & 2101.420 & -0.131 & 5.038E-20 & 5.580E-20 & -10.74 \\[-1.5mm]
$F_1$ & 14 & 4 & $F_2$ & 15 & 4 & 2101.294 & 2101.420 & -0.126 & 9.089E-21 & 7.565E-21 &  16.77 \\[-1.5mm]
$F_2$ & 14 & 2 & $F_1$ & 15 & 2 & 2101.310 & 2101.440 & -0.130 & 5.713E-21 & 6.863E-21 & -20.13 \\[-1.5mm]
$F_1$ & 14 & 3 & $F_2$ & 15 & 4 & 2101.345 & 2101.472 & -0.127 & 1.368E-20 & 1.974E-20 & -44.37 \\[-1.5mm]
$F_2$ & 14 & 5 & $F_1$ & 15 & 4 & 2101.369 & 2101.496 & -0.127 & 2.202E-20 & 2.588E-20 & -17.52 \\[-1.5mm]
$A_2$ & 14 & 4 & $A_1$ & 15 & 4 & 2101.397 & 2101.523 & -0.126 & 3.615E-20 & 5.131E-20 & -41.95 \\[-1.5mm]
$E$   & 14 & 2 & $E$   & 15 & 1 & 2101.445 & 2101.569 & -0.124 & 2.451E-21 & 3.595E-21 & -46.67 \\[-1.5mm]
$A_2$ & 15 & 4 & $A_1$ & 16 & 0 & 2096.608 & 2096.799 & -0.191 & 2.530E-20 & 3.021E-20 & -19.42 \\[-1.5mm]
$E$   & 15 & 2 & $E$   & 16 & 1 & 2096.658 & 2096.850 & -0.192 & 9.113E-21 & 1.064E-20 & -16.71 \\[-1.5mm]
$F_1$ & 15 & 2 & $F_2$ & 16 & 3 & 2096.686 & 2096.897 & -0.211 & 1.131E-20 & 1.532E-20 & -35.38 \\[-1.5mm]
$E$   & 15 & 6 & $E$   & 16 & 7 & 2096.743 & 2096.963 & -0.220 & 8.317E-21 & 9.454E-21 & -13.67 \\[-1.5mm]
$F_1$ & 15 & 3 & $F_2$ & 16 & 1 & 2096.772 & 2096.994 & -0.222 & 9.262E-21 & 1.085E-20 & -17.14 \\[-1.5mm]
$F_2$ & 15 & 7 & $F_1$ & 16 & 2 & 2096.802 & 2097.017 & -0.215 & 9.495E-21 & 1.206E-20 & -26.98 \\
\hline\hline
\end{longtable*}

 Due to the computational demands of calculating higher rotational excitation (rovibrational matrices scale linearly with $J$), calculations were performed with $P_{\mathrm{max}}=10$. Convergence tests were carried out up to $J=6$ for $P_{\mathrm{max}}=12$. The corresponding transition frequencies showed a consistent correction of around $\Delta (P_{\mathrm{max}}\!=\!12)=-0.00185{\,}$cm$^{-1}$. This correction was applied to all computed frequencies listed in Table~\ref{tab:v3_intens}. For the corresponding intensities, the $1\leftarrow 2$ $(J\p\leftarrow J\pp)$ transitions possessed a convergence correction of the order $10^{-24}$. The magnitude of this correction showed a linear relationship with increasing $J$, from which we estimate that for the $15\leftarrow 16$ transitions the correction would be of the order $10^{-22}$. The respective intensities therefore have an error of at most $1\%$. We are confident that the results in Table~\ref{tab:v3_intens} are sufficiently converged to reliably evaluate the DMS and PES.
 
 Around one third of the calculated absolute line intensities are within the estimated experimental measurement accuracy of $10\%$. However, as is best seen by the residuals plotted in Fig.~\eqref{fig:v3_band}, nearly all of the computed line intensities are larger than the corresponding experimental values. We suspect this is due to the electronic structure calculations and the use of only a triple-zeta basis set, aug-cc-pVTZ(+d for Si), to generate the DMS. A larger (augmented) correlation consistent basis set and possibly the inclusion of additional higher-level corrections (such as those incorporated for the PES) would most likely reduce the strength of computed line intensities. Despite this, Fig.~\eqref{fig:v3_band} shows that the $\nu_3$ band is well reproduced. Computed frequencies are on average larger by $0.1\!-\!0.2{\,}$cm$^{-1}$ across all transitions. This more or less systematic error can be attributed to the minor empirical refinement of the equilibrium Si-H bond length.

\subsection{Overview of rotation-vibration spectrum}
\label{sec:overview}

\begin{figure*}
\includegraphics{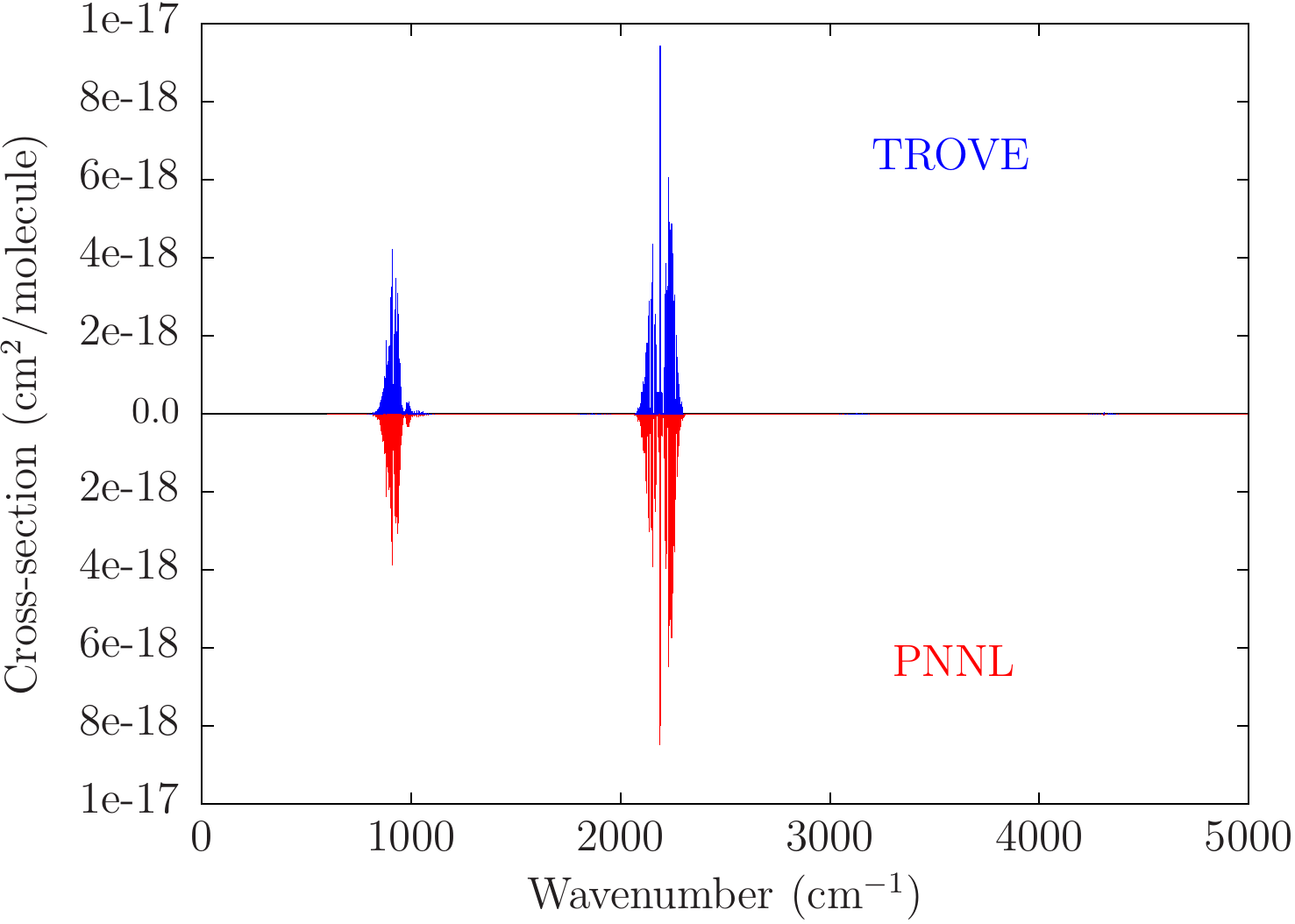}
\caption{\label{fig:pnnl}Overview of simulated $^{28}$SiH$_4$ rotation-vibration spectrum up to $J=20$. Note that the experimental PNNL spectrum~\cite{PNNL} is composed of $^{28}$SiH$_4$ ($92.2\%$), $^{29}$SiH$_4$ ($4.7\%$), and $^{30}$SiH$_4$ ($3.1\%$) (see text).}
\end{figure*}

 As a final test of the PES and DMS, in Fig.~\eqref{fig:pnnl} we have simulated the rotation-vibration spectrum of $^{28}$SiH$_4$ for transitions up to $J=20$ at $296{\,}$K. A polyad number of $P_{\mathrm{max}}=10$ was employed. Transition frequencies and corresponding intensities were calculated for a $5000{\,}$cm$^{-1}$ frequency window with a lower state energy threshold of $5000{\,}$cm$^{-1}$. To simulate the spectrum a Gaussian profile with a half width at half maximum of $0.135{\,}$cm$^{-1}$ was chosen as this appears to closely match the line shape used by the PNNL spectral library.~\cite{PNNL} The experimental PNNL silane spectrum, also shown in Fig.~\eqref{fig:pnnl}, is at a resolution of around $0.06{\,}$cm$^{-1}$. It was measured at a temperature of 25$\,^{\circ}$C with the dataset subsequently re-normalized to 22.84$\,^{\circ}$C ($296{\,}$K). Note that the PNNL spectrum is of electronics grade silane gas which is composed of $^{28}$SiH$_4$ ($92.2\%$), $^{29}$SiH$_4$ ($4.7\%$), and $^{30}$SiH$_4$ ($3.1\%$). We have therefore scaled the TROVE computed $^{28}$SiH$_4$ cross-sections by $0.922$ to provide a reliable comparison.

 The computed TROVE intensities are marginally stronger but overall there is good agreement with the experimental PNNL results. Even with $P_{\mathrm{max}}=10$ which does not give fully converged transition frequencies both band shape and position appear reliable. Of course there are shortcomings in our simulations which we will now discuss. 
 
 Some of the band structure is undoubtedly lost as we have not considered $^{29}$SiH$_4$ or $^{30}$SiH$_4$, and by only computing transitions up to $J=20$ the spectrum is unlikely to be complete at room temperature. There may also be minor errors arising from the use of a Gaussian profile to model the line shape. More desirable would be to fit a Voigt profile which incorporates instrumental factors. The largest source of error, as discussed before, is likely to be the electronic structure calculations. For the purposes of modelling exoplanet atmospheres however, we expect that the level of theory employed to compute the DMS is sufficient. The features of the SiH$_4$ spectrum are clear and identifiable as seen in Fig.~\eqref{fig:pnnl}.
 
\begin{figure*}
\includegraphics{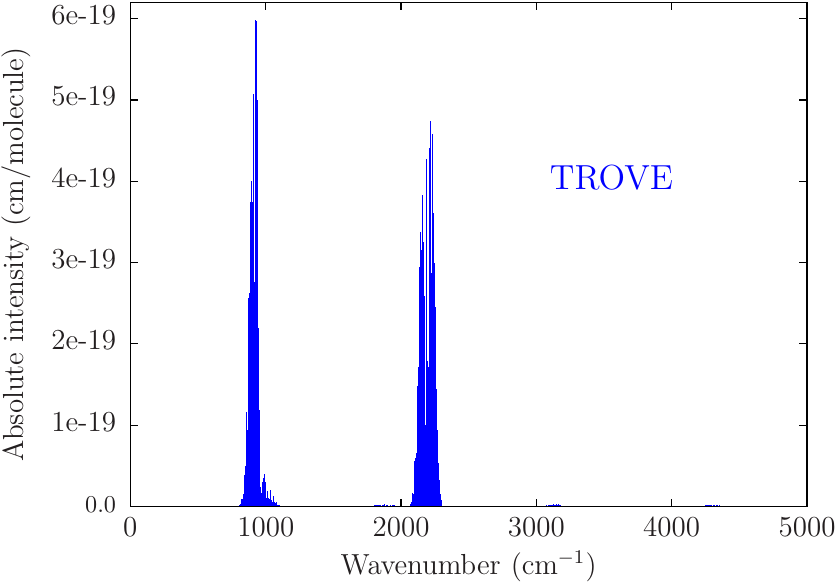}
\caption{\label{fig:stick_5000}Overview of absolute line intensities of $^{28}$SiH$_4$ up to $J=20$.}
\end{figure*}

 Note that in Fig.~\eqref{fig:pnnl} the $\nu_3$ ($2189.19{\,}$cm$^{-1}$) band is stronger than the $\nu_4$ ($913.47{\,}$cm$^{-1}$) band. This is contrast to the vibrational transition moments where $\mu_{\nu_4}>\mu_{\nu_3}$. If however we plot absolute line intensities up to $J=20$ as shown in Fig.~\eqref{fig:stick_5000}, the $\nu_4$ band is indeed stronger than the $\nu_3$ band. The behaviour displayed in Fig.~\eqref{fig:pnnl} is caused by the use of a line profile to model the spectrum.
 
\section{Conclusions}
\label{sec:conc}

 High-level \textit{ab initio} theory has been used to generate global potential energy and dipole moment surfaces for silane. The quality of the PES is reflected by the achievement of sub-wavenumber accuracy for all four fundamental frequencies. Combination and overtone bands are also consistently reproduced which confirms that the level of \textit{ab initio} theory used to generate the PES is adequate. Minor empirical refinement of the equilibrium geometry of SiH$_4$ produced an Si-H bond length in excellent agreement with previous experimental and theoretical results. The rotational structure of vibrational bands was improved as a result of the refinement. Ultimately though, to achieve sub-wavenumber accuracy for all rotation-vibration energy levels a rigorous empirical refinement of the PES is necessary.~\cite{YuBaTe11.NH3}
 
 A new \textit{ab initio} DMS has been computed and utilized to simulate the infrared spectrum of SiH$_4$. Absolute line intensities are marginally overestimated and we suspect this behaviour can be resolved by using a larger basis set for the electronic structure calculations when computing the DMS. Overall however, band shape and structure across the spectrum display good agreement with experiment. The PES and DMS presented in this work will be used to compute a comprehensive rovibrational line list applicable for elevated temperatures as part of the ExoMol project.~\cite{ExoMol2012,ExoMol2016}

\begin{acknowledgments}
This work was supported by ERC Advanced Investigator Project 267219, and FP7-MC-IEF project 629237.
\end{acknowledgments}


\begin{thebibliography}{91}%
\makeatletter
\providecommand \@ifxundefined [1]{%
 \@ifx{#1\undefined}
}%
\providecommand \@ifnum [1]{%
 \ifnum #1\expandafter \@firstoftwo
 \else \expandafter \@secondoftwo
 \fi
}%
\providecommand \@ifx [1]{%
 \ifx #1\expandafter \@firstoftwo
 \else \expandafter \@secondoftwo
 \fi
}%
\providecommand \natexlab [1]{#1}%
\providecommand \enquote  [1]{``#1''}%
\providecommand \bibnamefont  [1]{#1}%
\providecommand \bibfnamefont [1]{#1}%
\providecommand \citenamefont [1]{#1}%
\providecommand \href@noop [0]{\@secondoftwo}%
\providecommand \href [0]{\begingroup \@sanitize@url \@href}%
\providecommand \@href[1]{\@@startlink{#1}\@@href}%
\providecommand \@@href[1]{\endgroup#1\@@endlink}%
\providecommand \@sanitize@url [0]{\catcode `\\12\catcode `\$12\catcode
  `\&12\catcode `\#12\catcode `\^12\catcode `\_12\catcode `\%12\relax}%
\providecommand \@@startlink[1]{}%
\providecommand \@@endlink[0]{}%
\providecommand \url  [0]{\begingroup\@sanitize@url \@url }%
\providecommand \@url [1]{\endgroup\@href {#1}{\urlprefix }}%
\providecommand \urlprefix  [0]{URL }%
\providecommand \Eprint [0]{\href }%
\providecommand \doibase [0]{http://dx.doi.org/}%
\providecommand \selectlanguage [0]{\@gobble}%
\providecommand \bibinfo  [0]{\@secondoftwo}%
\providecommand \bibfield  [0]{\@secondoftwo}%
\providecommand \translation [1]{[#1]}%
\providecommand \BibitemOpen [0]{}%
\providecommand \bibitemStop [0]{}%
\providecommand \bibitemNoStop [0]{.\EOS\space}%
\providecommand \EOS [0]{\spacefactor3000\relax}%
\providecommand \BibitemShut  [1]{\csname bibitem#1\endcsname}%
\let\auto@bib@innerbib\@empty
\bibitem [{\citenamefont {Steward}\ and\ \citenamefont
  {Nielsen}(1934)}]{34StNixx.SiH4}%
  \BibitemOpen
  \bibfield  {author} {\bibinfo {author} {\bibfnamefont {W.~B.}\ \bibnamefont
  {Steward}}\ and\ \bibinfo {author} {\bibfnamefont {H.~H.}\ \bibnamefont
  {Nielsen}},\ }\href@noop {} {\bibfield  {journal} {\bibinfo  {journal} {J.
  Chem. Phys.}\ }\textbf {\bibinfo {volume} {2}},\ \bibinfo {pages} {712}
  (\bibinfo {year} {1934})}\BibitemShut {NoStop}%
\bibitem [{\citenamefont {Steward}\ and\ \citenamefont
  {Nielsen}(1935)}]{35StNixx.SiH4}%
  \BibitemOpen
  \bibfield  {author} {\bibinfo {author} {\bibfnamefont {W.~B.}\ \bibnamefont
  {Steward}}\ and\ \bibinfo {author} {\bibfnamefont {H.~H.}\ \bibnamefont
  {Nielsen}},\ }\href {\doibase 10.1103/PhysRev.47.828} {\bibfield  {journal}
  {\bibinfo  {journal} {Phys. Rev.}\ }\textbf {\bibinfo {volume} {47}},\
  \bibinfo {pages} {828} (\bibinfo {year} {1935})}\BibitemShut {NoStop}%
\bibitem [{\citenamefont {Goldhaber}\ and\ \citenamefont
  {Betz}(1984)}]{84GoBexx.SiH4}%
  \BibitemOpen
  \bibfield  {author} {\bibinfo {author} {\bibfnamefont {D.~M.}\ \bibnamefont
  {Goldhaber}}\ and\ \bibinfo {author} {\bibfnamefont {A.~L.}\ \bibnamefont
  {Betz}},\ }\href {\doibase 10.1086/184255} {\bibfield  {journal} {\bibinfo
  {journal} {Astrophys. J.}\ }\textbf {\bibinfo {volume} {279}},\ \bibinfo
  {pages} {L55} (\bibinfo {year} {1984})}\BibitemShut {NoStop}%
\bibitem [{\citenamefont {Keady}\ and\ \citenamefont
  {Ridgway}(1993)}]{93KeRixx.SiH4}%
  \BibitemOpen
  \bibfield  {author} {\bibinfo {author} {\bibfnamefont {J.~J.}\ \bibnamefont
  {Keady}}\ and\ \bibinfo {author} {\bibfnamefont {S.~T.}\ \bibnamefont
  {Ridgway}},\ }\href {\doibase 10.1086/172431} {\bibfield  {journal} {\bibinfo
   {journal} {Astrophys. J.}\ }\textbf {\bibinfo {volume} {406}},\ \bibinfo
  {pages} {199} (\bibinfo {year} {1993})}\BibitemShut {NoStop}%
\bibitem [{\citenamefont {Monnier}\ \emph {et~al.}(2000)\citenamefont
  {Monnier}, \citenamefont {Danchi}, \citenamefont {Hale}, \citenamefont
  {Tuthill},\ and\ \citenamefont {Townes}}]{00MoDaHa.SiH4}%
  \BibitemOpen
  \bibfield  {author} {\bibinfo {author} {\bibfnamefont {J.~D.}\ \bibnamefont
  {Monnier}}, \bibinfo {author} {\bibfnamefont {W.~C.}\ \bibnamefont {Danchi}},
  \bibinfo {author} {\bibfnamefont {D.~S.}\ \bibnamefont {Hale}}, \bibinfo
  {author} {\bibfnamefont {P.~G.}\ \bibnamefont {Tuthill}}, \ and\ \bibinfo
  {author} {\bibfnamefont {C.~H.}\ \bibnamefont {Townes}},\ }\href {\doibase
  10.1086/317127} {\bibfield  {journal} {\bibinfo  {journal} {Astrophys. J.}\
  }\textbf {\bibinfo {volume} {543}},\ \bibinfo {pages} {868} (\bibinfo {year}
  {2000})}\BibitemShut {NoStop}%
\bibitem [{\citenamefont {Treffers}\ \emph {et~al.}(1978)\citenamefont
  {Treffers}, \citenamefont {Larson}, \citenamefont {Fink},\ and\ \citenamefont
  {Gautier}}]{78TrLaFi.SiH4}%
  \BibitemOpen
  \bibfield  {author} {\bibinfo {author} {\bibfnamefont {R.~R.}\ \bibnamefont
  {Treffers}}, \bibinfo {author} {\bibfnamefont {H.~P.}\ \bibnamefont
  {Larson}}, \bibinfo {author} {\bibfnamefont {U.}~\bibnamefont {Fink}}, \ and\
  \bibinfo {author} {\bibfnamefont {T.~N.}\ \bibnamefont {Gautier}},\ }\href
  {\doibase 10.1016/0019-1035(78)90171-9} {\bibfield  {journal} {\bibinfo
  {journal} {Icarus}\ }\textbf {\bibinfo {volume} {34}},\ \bibinfo {pages}
  {331} (\bibinfo {year} {1978})}\BibitemShut {NoStop}%
\bibitem [{\citenamefont {Larson}\ \emph {et~al.}(1980)\citenamefont {Larson},
  \citenamefont {Fink}, \citenamefont {Smith},\ and\ \citenamefont
  {Davis}}]{80LaFiSm.SiH4}%
  \BibitemOpen
  \bibfield  {author} {\bibinfo {author} {\bibfnamefont {H.~P.}\ \bibnamefont
  {Larson}}, \bibinfo {author} {\bibfnamefont {U.}~\bibnamefont {Fink}},
  \bibinfo {author} {\bibfnamefont {H.~A.}\ \bibnamefont {Smith}}, \ and\
  \bibinfo {author} {\bibfnamefont {D.~S.}\ \bibnamefont {Davis}},\ }\href
  {\doibase 10.1086/158236} {\bibfield  {journal} {\bibinfo  {journal}
  {Astrophys. J.}\ }\textbf {\bibinfo {volume} {240}},\ \bibinfo {pages} {327}
  (\bibinfo {year} {1980})}\BibitemShut {NoStop}%
\bibitem [{\citenamefont {Martin}, \citenamefont {Baldridge},\ and\
  \citenamefont {Lee}(1999)}]{99MaBaLe.SiH4}%
  \BibitemOpen
  \bibfield  {author} {\bibinfo {author} {\bibfnamefont {J.~M.~L.}\
  \bibnamefont {Martin}}, \bibinfo {author} {\bibfnamefont {K.~K.}\
  \bibnamefont {Baldridge}}, \ and\ \bibinfo {author} {\bibfnamefont {T.~J.}\
  \bibnamefont {Lee}},\ }\href@noop {} {\bibfield  {journal} {\bibinfo
  {journal} {Mol. Phys.}\ }\textbf {\bibinfo {volume} {97}},\ \bibinfo {pages}
  {945} (\bibinfo {year} {1999})}\BibitemShut {NoStop}%
\bibitem [{\citenamefont {Dunning~Jr.}(1989)}]{Dunning89}%
  \BibitemOpen
  \bibfield  {author} {\bibinfo {author} {\bibfnamefont {T.~H.}\ \bibnamefont
  {Dunning~Jr.}},\ }\href {\doibase 10.1063/1.456153} {\bibfield  {journal}
  {\bibinfo  {journal} {J. Chem. Phys.}\ }\textbf {\bibinfo {volume} {90}},\
  \bibinfo {pages} {1007} (\bibinfo {year} {1989})}\BibitemShut {NoStop}%
\bibitem [{\citenamefont {Martin}\ and\ \citenamefont
  {Uzan}(1998)}]{Martin:1998}%
  \BibitemOpen
  \bibfield  {author} {\bibinfo {author} {\bibfnamefont {J.~M.~L.}\
  \bibnamefont {Martin}}\ and\ \bibinfo {author} {\bibfnamefont
  {O.}~\bibnamefont {Uzan}},\ }\href {\doibase 10.1016/S0009-2614(97)01128-7}
  {\bibfield  {journal} {\bibinfo  {journal} {Chem. Phys. Lett.}\ }\textbf
  {\bibinfo {volume} {282}},\ \bibinfo {pages} {16} (\bibinfo {year}
  {1998})}\BibitemShut {NoStop}%
\bibitem [{\citenamefont {Wang}\ and\ \citenamefont
  {Sibert}(2000)}]{00WaSixx.SiH4}%
  \BibitemOpen
  \bibfield  {author} {\bibinfo {author} {\bibfnamefont {X.~G.}\ \bibnamefont
  {Wang}}\ and\ \bibinfo {author} {\bibfnamefont {E.~L.}\ \bibnamefont
  {Sibert}},\ }\href {\doibase 10.1063/1.1290027} {\bibfield  {journal}
  {\bibinfo  {journal} {J. Chem. Phys.}\ }\textbf {\bibinfo {volume} {113}},\
  \bibinfo {pages} {5384} (\bibinfo {year} {2000})}\BibitemShut {NoStop}%
\bibitem [{\citenamefont {Hou}, \citenamefont {Borondo},\ and\ \citenamefont
  {Benito}(2001)}]{01HoBoBe.SiH4}%
  \BibitemOpen
  \bibfield  {author} {\bibinfo {author} {\bibfnamefont {X.~W.}\ \bibnamefont
  {Hou}}, \bibinfo {author} {\bibfnamefont {F.}~\bibnamefont {Borondo}}, \ and\
  \bibinfo {author} {\bibfnamefont {R.~M.}\ \bibnamefont {Benito}},\ }\href
  {\doibase 10.1016/S0009-2614(01)00825-9} {\bibfield  {journal} {\bibinfo
  {journal} {Chem. Phys. Lett.}\ }\textbf {\bibinfo {volume} {344}},\ \bibinfo
  {pages} {421} (\bibinfo {year} {2001})}\BibitemShut {NoStop}%
\bibitem [{\citenamefont {Halonen}\ and\ \citenamefont
  {Child}(1982)}]{82HaChxx.SiH4}%
  \BibitemOpen
  \bibfield  {author} {\bibinfo {author} {\bibfnamefont {L.}~\bibnamefont
  {Halonen}}\ and\ \bibinfo {author} {\bibfnamefont {M.~S.}\ \bibnamefont
  {Child}},\ }\href {\doibase 10.1080/00268978200101231} {\bibfield  {journal}
  {\bibinfo  {journal} {Mol. Phys.}\ }\textbf {\bibinfo {volume} {46}},\
  \bibinfo {pages} {239} (\bibinfo {year} {1982})}\BibitemShut {NoStop}%
\bibitem [{\citenamefont {Halonen}\ and\ \citenamefont
  {Child}(1988)}]{88HaChxx.SiH4}%
  \BibitemOpen
  \bibfield  {author} {\bibinfo {author} {\bibfnamefont {L.}~\bibnamefont
  {Halonen}}\ and\ \bibinfo {author} {\bibfnamefont {M.~S.}\ \bibnamefont
  {Child}},\ }\href {\doibase 10.1016/0010-4655(88)90070-7} {\bibfield
  {journal} {\bibinfo  {journal} {Comput. Phys. Commun.}\ }\textbf {\bibinfo
  {volume} {51}},\ \bibinfo {pages} {173} (\bibinfo {year} {1988})}\BibitemShut
  {NoStop}%
\bibitem [{\citenamefont {Permogorov}\ and\ \citenamefont
  {Campargue}(1997)}]{97PeCaxx.SiH4}%
  \BibitemOpen
  \bibfield  {author} {\bibinfo {author} {\bibfnamefont {D.}~\bibnamefont
  {Permogorov}}\ and\ \bibinfo {author} {\bibfnamefont {A.}~\bibnamefont
  {Campargue}},\ }\href {\doibase 10.1080/002689797170680} {\bibfield
  {journal} {\bibinfo  {journal} {Mol. Phys.}\ }\textbf {\bibinfo {volume}
  {92}},\ \bibinfo {pages} {117} (\bibinfo {year} {1997})}\BibitemShut
  {NoStop}%
\bibitem [{\citenamefont {Xie}\ and\ \citenamefont
  {Tennyson}(2002)}]{02XiTexx.SiH4}%
  \BibitemOpen
  \bibfield  {author} {\bibinfo {author} {\bibfnamefont {J.~K.}\ \bibnamefont
  {Xie}}\ and\ \bibinfo {author} {\bibfnamefont {J.}~\bibnamefont {Tennyson}},\
  }\href {\doibase 10.1080/00268970210126628} {\bibfield  {journal} {\bibinfo
  {journal} {Mol. Phys.}\ }\textbf {\bibinfo {volume} {100}},\ \bibinfo {pages}
  {1615} (\bibinfo {year} {2002})}\BibitemShut {NoStop}%
\bibitem [{\citenamefont {Lin}\ \emph {et~al.}(1998)\citenamefont {Lin},
  \citenamefont {Wang}, \citenamefont {Chen}, \citenamefont {Wang},
  \citenamefont {Zhou},\ and\ \citenamefont {Zhu}}]{98LiWaCh.SiH4}%
  \BibitemOpen
  \bibfield  {author} {\bibinfo {author} {\bibfnamefont {H.}~\bibnamefont
  {Lin}}, \bibinfo {author} {\bibfnamefont {D.}~\bibnamefont {Wang}}, \bibinfo
  {author} {\bibfnamefont {X.~Y.}\ \bibnamefont {Chen}}, \bibinfo {author}
  {\bibfnamefont {X.~G.}\ \bibnamefont {Wang}}, \bibinfo {author}
  {\bibfnamefont {Z.~P.}\ \bibnamefont {Zhou}}, \ and\ \bibinfo {author}
  {\bibfnamefont {Q.~S.}\ \bibnamefont {Zhu}},\ }\href {\doibase
  10.1006/jmsp.1998.7673} {\bibfield  {journal} {\bibinfo  {journal} {J. Mol.
  Spectrosc.}\ }\textbf {\bibinfo {volume} {192}},\ \bibinfo {pages} {249}
  (\bibinfo {year} {1998})}\BibitemShut {NoStop}%
\bibitem [{\citenamefont {Lin}, \citenamefont {Yuan},\ and\ \citenamefont
  {Zhu}(1999)}]{99LiYuZh.SiH4}%
  \BibitemOpen
  \bibfield  {author} {\bibinfo {author} {\bibfnamefont {H.}~\bibnamefont
  {Lin}}, \bibinfo {author} {\bibfnamefont {L.~F.}\ \bibnamefont {Yuan}}, \
  and\ \bibinfo {author} {\bibfnamefont {Q.~S.}\ \bibnamefont {Zhu}},\ }\href
  {\doibase 10.1016/S0009-2614(99)00567-9} {\bibfield  {journal} {\bibinfo
  {journal} {Chem. Phys. Lett.}\ }\textbf {\bibinfo {volume} {308}},\ \bibinfo
  {pages} {137} (\bibinfo {year} {1999})}\BibitemShut {NoStop}%
\bibitem [{\citenamefont {Lin}\ \emph {et~al.}(2001{\natexlab{a}})\citenamefont
  {Lin}, \citenamefont {He}, \citenamefont {Wang}, \citenamefont {Yuan},
  \citenamefont {B\"{u}rger}, \citenamefont {D'Eu}, \citenamefont {Reuter},\
  and\ \citenamefont {Thiel}}]{01LiHeWa.SiH4}%
  \BibitemOpen
  \bibfield  {author} {\bibinfo {author} {\bibfnamefont {H.}~\bibnamefont
  {Lin}}, \bibinfo {author} {\bibfnamefont {S.~G.}\ \bibnamefont {He}},
  \bibinfo {author} {\bibfnamefont {X.~G.}\ \bibnamefont {Wang}}, \bibinfo
  {author} {\bibfnamefont {L.~F.}\ \bibnamefont {Yuan}}, \bibinfo {author}
  {\bibfnamefont {H.}~\bibnamefont {B\"{u}rger}}, \bibinfo {author}
  {\bibfnamefont {J.~F.}\ \bibnamefont {D'Eu}}, \bibinfo {author}
  {\bibfnamefont {N.}~\bibnamefont {Reuter}}, \ and\ \bibinfo {author}
  {\bibfnamefont {W.}~\bibnamefont {Thiel}},\ }\href {\doibase
  10.1039/b104487g} {\bibfield  {journal} {\bibinfo  {journal} {Phys. Chem.
  Chem. Phys.}\ }\textbf {\bibinfo {volume} {3}},\ \bibinfo {pages} {3506}
  (\bibinfo {year} {2001}{\natexlab{a}})}\BibitemShut {NoStop}%
\bibitem [{\citenamefont {He}\ \emph {et~al.}(2002)\citenamefont {He},
  \citenamefont {Liu}, \citenamefont {Lin}, \citenamefont {Hu}, \citenamefont
  {Zheng}, \citenamefont {Hao},\ and\ \citenamefont {Zhu}}]{02HeLiLi.SiH4}%
  \BibitemOpen
  \bibfield  {author} {\bibinfo {author} {\bibfnamefont {S.~G.}\ \bibnamefont
  {He}}, \bibinfo {author} {\bibfnamefont {A.~W.}\ \bibnamefont {Liu}},
  \bibinfo {author} {\bibfnamefont {H.}~\bibnamefont {Lin}}, \bibinfo {author}
  {\bibfnamefont {S.~M.}\ \bibnamefont {Hu}}, \bibinfo {author} {\bibfnamefont
  {J.~J.}\ \bibnamefont {Zheng}}, \bibinfo {author} {\bibfnamefont {L.~Y.}\
  \bibnamefont {Hao}}, \ and\ \bibinfo {author} {\bibfnamefont {Q.~S.}\
  \bibnamefont {Zhu}},\ }\href {\doibase 10.1063/1.1520130} {\bibfield
  {journal} {\bibinfo  {journal} {J. Chem. Phys.}\ }\textbf {\bibinfo {volume}
  {117}},\ \bibinfo {pages} {10073} (\bibinfo {year} {2002})}\BibitemShut
  {NoStop}%
\bibitem [{\citenamefont {Zhu}\ \emph {et~al.}(1989)\citenamefont {Zhu},
  \citenamefont {Zhang}, \citenamefont {Ma},\ and\ \citenamefont
  {Qian}}]{89ZhZhMa.SiH4}%
  \BibitemOpen
  \bibfield  {author} {\bibinfo {author} {\bibfnamefont {Q.~S.}\ \bibnamefont
  {Zhu}}, \bibinfo {author} {\bibfnamefont {B.~S.}\ \bibnamefont {Zhang}},
  \bibinfo {author} {\bibfnamefont {Y.~R.}\ \bibnamefont {Ma}}, \ and\ \bibinfo
  {author} {\bibfnamefont {H.~B.}\ \bibnamefont {Qian}},\ }\href {\doibase
  10.1016/0009-2614(89)85265-0} {\bibfield  {journal} {\bibinfo  {journal}
  {Chem. Phys. Lett.}\ }\textbf {\bibinfo {volume} {164}},\ \bibinfo {pages}
  {596} (\bibinfo {year} {1989})}\BibitemShut {NoStop}%
\bibitem [{\citenamefont {Zhu}\ \emph {et~al.}(1990{\natexlab{a}})\citenamefont
  {Zhu}, \citenamefont {Zhang}, \citenamefont {Ma},\ and\ \citenamefont
  {Qian}}]{90ZhZhMa.SiH4}%
  \BibitemOpen
  \bibfield  {author} {\bibinfo {author} {\bibfnamefont {Q.~S.}\ \bibnamefont
  {Zhu}}, \bibinfo {author} {\bibfnamefont {B.~S.}\ \bibnamefont {Zhang}},
  \bibinfo {author} {\bibfnamefont {Y.~R.}\ \bibnamefont {Ma}}, \ and\ \bibinfo
  {author} {\bibfnamefont {H.~B.}\ \bibnamefont {Qian}},\ }\href {\doibase
  10.1016/0584-8539(90)80198-8} {\bibfield  {journal} {\bibinfo  {journal}
  {Spectrochim. Acta A}\ }\textbf {\bibinfo {volume} {46}},\ \bibinfo {pages}
  {1217} (\bibinfo {year} {1990}{\natexlab{a}})}\BibitemShut {NoStop}%
\bibitem [{\citenamefont {Zhu}\ \emph {et~al.}(1990{\natexlab{b}})\citenamefont
  {Zhu}, \citenamefont {Ma}, \citenamefont {Zhang}, \citenamefont {Ma},\ and\
  \citenamefont {Qian}}]{90ZhMaZh.SiH4}%
  \BibitemOpen
  \bibfield  {author} {\bibinfo {author} {\bibfnamefont {Q.~S.}\ \bibnamefont
  {Zhu}}, \bibinfo {author} {\bibfnamefont {H.}~\bibnamefont {Ma}}, \bibinfo
  {author} {\bibfnamefont {B.~S.}\ \bibnamefont {Zhang}}, \bibinfo {author}
  {\bibfnamefont {Y.~R.}\ \bibnamefont {Ma}}, \ and\ \bibinfo {author}
  {\bibfnamefont {H.~B.}\ \bibnamefont {Qian}},\ }\href {\doibase
  10.1016/0584-8539(90)80137-N} {\bibfield  {journal} {\bibinfo  {journal}
  {Spectrochim. Acta A}\ }\textbf {\bibinfo {volume} {46}},\ \bibinfo {pages}
  {1323} (\bibinfo {year} {1990}{\natexlab{b}})}\BibitemShut {NoStop}%
\bibitem [{\citenamefont {Zhu}\ \emph {et~al.}(1991)\citenamefont {Zhu},
  \citenamefont {Qian}, \citenamefont {Ma},\ and\ \citenamefont
  {Halonen}}]{91ZhQiMa.SiH4}%
  \BibitemOpen
  \bibfield  {author} {\bibinfo {author} {\bibfnamefont {Q.~S.}\ \bibnamefont
  {Zhu}}, \bibinfo {author} {\bibfnamefont {H.~B.}\ \bibnamefont {Qian}},
  \bibinfo {author} {\bibfnamefont {H.}~\bibnamefont {Ma}}, \ and\ \bibinfo
  {author} {\bibfnamefont {L.}~\bibnamefont {Halonen}},\ }\href {\doibase
  10.1016/0009-2614(91)85026-S} {\bibfield  {journal} {\bibinfo  {journal}
  {Chem. Phys. Lett.}\ }\textbf {\bibinfo {volume} {177}},\ \bibinfo {pages}
  {261} (\bibinfo {year} {1991})}\BibitemShut {NoStop}%
\bibitem [{\citenamefont {Sun}\ \emph {et~al.}(1995)\citenamefont {Sun},
  \citenamefont {Wang}, \citenamefont {Zhu}, \citenamefont {Pierre},\ and\
  \citenamefont {Pierre}}]{95SuWaZh.SiH4}%
  \BibitemOpen
  \bibfield  {author} {\bibinfo {author} {\bibfnamefont {F.~G.}\ \bibnamefont
  {Sun}}, \bibinfo {author} {\bibfnamefont {X.~G.}\ \bibnamefont {Wang}},
  \bibinfo {author} {\bibfnamefont {Q.~S.}\ \bibnamefont {Zhu}}, \bibinfo
  {author} {\bibfnamefont {C.}~\bibnamefont {Pierre}}, \ and\ \bibinfo {author}
  {\bibfnamefont {G.}~\bibnamefont {Pierre}},\ }\href {\doibase
  10.1016/0009-2614(95)00475-J} {\bibfield  {journal} {\bibinfo  {journal}
  {Chem. Phys. Lett.}\ }\textbf {\bibinfo {volume} {239}},\ \bibinfo {pages}
  {373} (\bibinfo {year} {1995})}\BibitemShut {NoStop}%
\bibitem [{\citenamefont {Zhu}, \citenamefont {Campargue},\ and\ \citenamefont
  {Stoeckel}(1994)}]{94ZhCaSt.SiH4}%
  \BibitemOpen
  \bibfield  {author} {\bibinfo {author} {\bibfnamefont {Q.~S.}\ \bibnamefont
  {Zhu}}, \bibinfo {author} {\bibfnamefont {A.}~\bibnamefont {Campargue}}, \
  and\ \bibinfo {author} {\bibfnamefont {F.}~\bibnamefont {Stoeckel}},\ }\href
  {\doibase 10.1016/0584-8539(94)80001-4} {\bibfield  {journal} {\bibinfo
  {journal} {Spectrochim. Acta A}\ }\textbf {\bibinfo {volume} {50}},\ \bibinfo
  {pages} {663} (\bibinfo {year} {1994})}\BibitemShut {NoStop}%
\bibitem [{\citenamefont {Lin}\ \emph {et~al.}(2001{\natexlab{b}})\citenamefont
  {Lin}, \citenamefont {B\"{u}rger}, \citenamefont {He}, \citenamefont {Yuan},
  \citenamefont {Breidung},\ and\ \citenamefont {Thiel}}]{01LiBuHe.SiH4}%
  \BibitemOpen
  \bibfield  {author} {\bibinfo {author} {\bibfnamefont {H.}~\bibnamefont
  {Lin}}, \bibinfo {author} {\bibfnamefont {H.}~\bibnamefont {B\"{u}rger}},
  \bibinfo {author} {\bibfnamefont {S.~G.}\ \bibnamefont {He}}, \bibinfo
  {author} {\bibfnamefont {L.~F.}\ \bibnamefont {Yuan}}, \bibinfo {author}
  {\bibfnamefont {J.}~\bibnamefont {Breidung}}, \ and\ \bibinfo {author}
  {\bibfnamefont {W.}~\bibnamefont {Thiel}},\ }\href {\doibase
  10.1021/jp010404t} {\bibfield  {journal} {\bibinfo  {journal} {J. Phys. Chem.
  A}\ }\textbf {\bibinfo {volume} {105}},\ \bibinfo {pages} {6065} (\bibinfo
  {year} {2001}{\natexlab{b}})}\BibitemShut {NoStop}%
\bibitem [{\citenamefont {Yurchenko}(2014)}]{13Yuxxxx.method}%
  \BibitemOpen
  \bibfield  {author} {\bibinfo {author} {\bibfnamefont {S.~N.}\ \bibnamefont
  {Yurchenko}},\ }in\ \href {\doibase 10.1039/9781849737241-00183} {\emph
  {\bibinfo {booktitle} {Chemical Modelling: Volume 10}}},\ Vol.~\bibinfo
  {volume} {10}\ (\bibinfo  {publisher} {The Royal Society of Chemistry},\
  \bibinfo {year} {2014})\ pp.\ \bibinfo {pages} {183--228}\BibitemShut
  {NoStop}%
\bibitem [{\citenamefont {Tennyson}\ and\ \citenamefont
  {Yurchenko}(2012)}]{ExoMol2012}%
  \BibitemOpen
  \bibfield  {author} {\bibinfo {author} {\bibfnamefont {J.}~\bibnamefont
  {Tennyson}}\ and\ \bibinfo {author} {\bibfnamefont {S.~N.}\ \bibnamefont
  {Yurchenko}},\ }\href {\doibase 10.1111/j.1365-2966.2012.21440.x} {\bibfield
  {journal} {\bibinfo  {journal} {Mon. Not. R. Astron. Soc.}\ }\textbf
  {\bibinfo {volume} {425}},\ \bibinfo {pages} {21} (\bibinfo {year}
  {2012})}\BibitemShut {NoStop}%
\bibitem [{\citenamefont {Seager}, \citenamefont {Bains},\ and\ \citenamefont
  {Hu}(2013)}]{13aSeBaHu.CH3Cl}%
  \BibitemOpen
  \bibfield  {author} {\bibinfo {author} {\bibfnamefont {S.}~\bibnamefont
  {Seager}}, \bibinfo {author} {\bibfnamefont {W.}~\bibnamefont {Bains}}, \
  and\ \bibinfo {author} {\bibfnamefont {R.}~\bibnamefont {Hu}},\ }\href
  {\doibase 10.1088/0004-637X/777/2/95} {\bibfield  {journal} {\bibinfo
  {journal} {Astrophys. J.}\ }\textbf {\bibinfo {volume} {777}},\ \bibinfo
  {pages} {95} (\bibinfo {year} {2013})}\BibitemShut {NoStop}%
\bibitem [{\citenamefont {Rothman}\ \emph {et~al.}(2013)\citenamefont
  {Rothman}, \citenamefont {Gordon}, \citenamefont {Babikov}, \citenamefont
  {Barbe}, \citenamefont {Benner}, \citenamefont {Bernath}, \citenamefont
  {Birk}, \citenamefont {Bizzocchi}, \citenamefont {Boudon}, \citenamefont
  {Brown}, \citenamefont {Campargue}, \citenamefont {Chance}, \citenamefont
  {Cohen}, \citenamefont {Coudert}, \citenamefont {Devi}, \citenamefont
  {Drouin}, \citenamefont {Fayt}, \citenamefont {Flaud}, \citenamefont
  {Gamache}, \citenamefont {Harrison}, \citenamefont {Hartmann}, \citenamefont
  {Hill}, \citenamefont {Hodges}, \citenamefont {Jacquemart}, \citenamefont
  {Jolly}, \citenamefont {Lamouroux}, \citenamefont {Roy}, \citenamefont {Li},
  \citenamefont {Long}, \citenamefont {Lyulin}, \citenamefont {Mackie},
  \citenamefont {Massie}, \citenamefont {Mikhailenko}, \citenamefont
  {M{\"{u}}ller}, \citenamefont {Naumenko}, \citenamefont {Nikitin},
  \citenamefont {Orphal}, \citenamefont {Perevalov}, \citenamefont {Perrin},
  \citenamefont {Polovtseva}, \citenamefont {Richard}, \citenamefont {Smith},
  \citenamefont {Starikova}, \citenamefont {Sung}, \citenamefont {Tashkun},
  \citenamefont {Tennyson}, \citenamefont {Toon}, \citenamefont {Tyuterev},\
  and\ \citenamefont {Wagner}}]{HITRAN}%
  \BibitemOpen
  \bibfield  {author} {\bibinfo {author} {\bibfnamefont {L.}~\bibnamefont
  {Rothman}}, \bibinfo {author} {\bibfnamefont {I.}~\bibnamefont {Gordon}},
  \bibinfo {author} {\bibfnamefont {Y.}~\bibnamefont {Babikov}}, \bibinfo
  {author} {\bibfnamefont {A.}~\bibnamefont {Barbe}}, \bibinfo {author}
  {\bibfnamefont {D.~C.}\ \bibnamefont {Benner}}, \bibinfo {author}
  {\bibfnamefont {P.}~\bibnamefont {Bernath}}, \bibinfo {author} {\bibfnamefont
  {M.}~\bibnamefont {Birk}}, \bibinfo {author} {\bibfnamefont {L.}~\bibnamefont
  {Bizzocchi}}, \bibinfo {author} {\bibfnamefont {V.}~\bibnamefont {Boudon}},
  \bibinfo {author} {\bibfnamefont {L.}~\bibnamefont {Brown}}, \bibinfo
  {author} {\bibfnamefont {A.}~\bibnamefont {Campargue}}, \bibinfo {author}
  {\bibfnamefont {K.}~\bibnamefont {Chance}}, \bibinfo {author} {\bibfnamefont
  {E.}~\bibnamefont {Cohen}}, \bibinfo {author} {\bibfnamefont
  {L.}~\bibnamefont {Coudert}}, \bibinfo {author} {\bibfnamefont
  {V.}~\bibnamefont {Devi}}, \bibinfo {author} {\bibfnamefont {B.}~\bibnamefont
  {Drouin}}, \bibinfo {author} {\bibfnamefont {A.}~\bibnamefont {Fayt}},
  \bibinfo {author} {\bibfnamefont {J.-M.}\ \bibnamefont {Flaud}}, \bibinfo
  {author} {\bibfnamefont {R.}~\bibnamefont {Gamache}}, \bibinfo {author}
  {\bibfnamefont {J.}~\bibnamefont {Harrison}}, \bibinfo {author}
  {\bibfnamefont {J.-M.}\ \bibnamefont {Hartmann}}, \bibinfo {author}
  {\bibfnamefont {C.}~\bibnamefont {Hill}}, \bibinfo {author} {\bibfnamefont
  {J.}~\bibnamefont {Hodges}}, \bibinfo {author} {\bibfnamefont
  {D.}~\bibnamefont {Jacquemart}}, \bibinfo {author} {\bibfnamefont
  {A.}~\bibnamefont {Jolly}}, \bibinfo {author} {\bibfnamefont
  {J.}~\bibnamefont {Lamouroux}}, \bibinfo {author} {\bibfnamefont {R.~L.}\
  \bibnamefont {Roy}}, \bibinfo {author} {\bibfnamefont {G.}~\bibnamefont
  {Li}}, \bibinfo {author} {\bibfnamefont {D.}~\bibnamefont {Long}}, \bibinfo
  {author} {\bibfnamefont {O.}~\bibnamefont {Lyulin}}, \bibinfo {author}
  {\bibfnamefont {C.}~\bibnamefont {Mackie}}, \bibinfo {author} {\bibfnamefont
  {S.}~\bibnamefont {Massie}}, \bibinfo {author} {\bibfnamefont
  {S.}~\bibnamefont {Mikhailenko}}, \bibinfo {author} {\bibfnamefont
  {H.}~\bibnamefont {M{\"{u}}ller}}, \bibinfo {author} {\bibfnamefont
  {O.}~\bibnamefont {Naumenko}}, \bibinfo {author} {\bibfnamefont
  {A.}~\bibnamefont {Nikitin}}, \bibinfo {author} {\bibfnamefont
  {J.}~\bibnamefont {Orphal}}, \bibinfo {author} {\bibfnamefont
  {V.}~\bibnamefont {Perevalov}}, \bibinfo {author} {\bibfnamefont
  {A.}~\bibnamefont {Perrin}}, \bibinfo {author} {\bibfnamefont
  {E.}~\bibnamefont {Polovtseva}}, \bibinfo {author} {\bibfnamefont
  {C.}~\bibnamefont {Richard}}, \bibinfo {author} {\bibfnamefont
  {M.}~\bibnamefont {Smith}}, \bibinfo {author} {\bibfnamefont
  {E.}~\bibnamefont {Starikova}}, \bibinfo {author} {\bibfnamefont
  {K.}~\bibnamefont {Sung}}, \bibinfo {author} {\bibfnamefont {S.}~\bibnamefont
  {Tashkun}}, \bibinfo {author} {\bibfnamefont {J.}~\bibnamefont {Tennyson}},
  \bibinfo {author} {\bibfnamefont {G.}~\bibnamefont {Toon}}, \bibinfo {author}
  {\bibfnamefont {V.}~\bibnamefont {Tyuterev}}, \ and\ \bibinfo {author}
  {\bibfnamefont {G.}~\bibnamefont {Wagner}},\ }\href {\doibase
  http://dx.doi.org/10.1016/j.jqsrt.2013.07.002} {\bibfield  {journal}
  {\bibinfo  {journal} {J. Quant. Spectrosc. Radiat. Transf.}\ }\textbf
  {\bibinfo {volume} {130}},\ \bibinfo {pages} {4} (\bibinfo {year}
  {2013})}\BibitemShut {NoStop}%
\bibitem [{\citenamefont {Jacquinet-Husson}\ \emph {et~al.}(2011)\citenamefont
  {Jacquinet-Husson}, \citenamefont {Crepeau}, \citenamefont {Armante},
  \citenamefont {Boutammine}, \citenamefont {Ch\'{e}din}, \citenamefont
  {Scott}, \citenamefont {Crevoisier}, \citenamefont {Capelle}, \citenamefont
  {Boone}, \citenamefont {Poulet-Crovisier}, \citenamefont {Barbe},
  \citenamefont {Campargue}, \citenamefont {Benner}, \citenamefont {Benilan},
  \citenamefont {B\'{e}zard}, \citenamefont {Boudon}, \citenamefont {Brown},
  \citenamefont {Coudert}, \citenamefont {Coustenis}, \citenamefont {Dana},
  \citenamefont {Devi}, \citenamefont {Fally}, \citenamefont {Fayt},
  \citenamefont {Flaud}, \citenamefont {Goldman}, \citenamefont {Herman},
  \citenamefont {Harris}, \citenamefont {Jacquemart}, \citenamefont {Jolly},
  \citenamefont {Kleiner}, \citenamefont {Kleinboehl}, \citenamefont
  {Kwabia-Tchana}, \citenamefont {Lavrentieva}, \citenamefont {Lacome},
  \citenamefont {Xu}, \citenamefont {Lyulin}, \citenamefont {Mandin},
  \citenamefont {Maki}, \citenamefont {Mikhailenko}, \citenamefont {Miller},
  \citenamefont {Mishina}, \citenamefont {Moazzen-Ahmadi}, \citenamefont
  {M\"{u}ller}, \citenamefont {Nikitin}, \citenamefont {Orphal}, \citenamefont
  {Perevalov}, \citenamefont {Perrin}, \citenamefont {Petkie}, \citenamefont
  {Predoi-Cross}, \citenamefont {Rinsland}, \citenamefont {Remedios},
  \citenamefont {Rotger}, \citenamefont {Smith}, \citenamefont {Sung},
  \citenamefont {Tashkun}, \citenamefont {Tennyson}, \citenamefont {Toth},
  \citenamefont {Vandaele},\ and\ \citenamefont {Vander~Auwera}}]{GEISA}%
  \BibitemOpen
  \bibfield  {author} {\bibinfo {author} {\bibfnamefont {N.}~\bibnamefont
  {Jacquinet-Husson}}, \bibinfo {author} {\bibfnamefont {L.}~\bibnamefont
  {Crepeau}}, \bibinfo {author} {\bibfnamefont {R.}~\bibnamefont {Armante}},
  \bibinfo {author} {\bibfnamefont {C.}~\bibnamefont {Boutammine}}, \bibinfo
  {author} {\bibfnamefont {A.}~\bibnamefont {Ch\'{e}din}}, \bibinfo {author}
  {\bibfnamefont {N.~A.}\ \bibnamefont {Scott}}, \bibinfo {author}
  {\bibfnamefont {C.}~\bibnamefont {Crevoisier}}, \bibinfo {author}
  {\bibfnamefont {V.}~\bibnamefont {Capelle}}, \bibinfo {author} {\bibfnamefont
  {C.}~\bibnamefont {Boone}}, \bibinfo {author} {\bibfnamefont
  {N.}~\bibnamefont {Poulet-Crovisier}}, \bibinfo {author} {\bibfnamefont
  {A.}~\bibnamefont {Barbe}}, \bibinfo {author} {\bibfnamefont
  {A.}~\bibnamefont {Campargue}}, \bibinfo {author} {\bibfnamefont {D.~C.}\
  \bibnamefont {Benner}}, \bibinfo {author} {\bibfnamefont {Y.}~\bibnamefont
  {Benilan}}, \bibinfo {author} {\bibfnamefont {B.}~\bibnamefont {B\'{e}zard}},
  \bibinfo {author} {\bibfnamefont {V.}~\bibnamefont {Boudon}}, \bibinfo
  {author} {\bibfnamefont {L.~R.}\ \bibnamefont {Brown}}, \bibinfo {author}
  {\bibfnamefont {L.~H.}\ \bibnamefont {Coudert}}, \bibinfo {author}
  {\bibfnamefont {A.}~\bibnamefont {Coustenis}}, \bibinfo {author}
  {\bibfnamefont {V.}~\bibnamefont {Dana}}, \bibinfo {author} {\bibfnamefont
  {V.~M.}\ \bibnamefont {Devi}}, \bibinfo {author} {\bibfnamefont
  {S.}~\bibnamefont {Fally}}, \bibinfo {author} {\bibfnamefont
  {A.}~\bibnamefont {Fayt}}, \bibinfo {author} {\bibfnamefont {J.~M.}\
  \bibnamefont {Flaud}}, \bibinfo {author} {\bibfnamefont {A.}~\bibnamefont
  {Goldman}}, \bibinfo {author} {\bibfnamefont {M.}~\bibnamefont {Herman}},
  \bibinfo {author} {\bibfnamefont {G.~J.}\ \bibnamefont {Harris}}, \bibinfo
  {author} {\bibfnamefont {D.}~\bibnamefont {Jacquemart}}, \bibinfo {author}
  {\bibfnamefont {A.}~\bibnamefont {Jolly}}, \bibinfo {author} {\bibfnamefont
  {I.}~\bibnamefont {Kleiner}}, \bibinfo {author} {\bibfnamefont
  {A.}~\bibnamefont {Kleinboehl}}, \bibinfo {author} {\bibfnamefont
  {F.}~\bibnamefont {Kwabia-Tchana}}, \bibinfo {author} {\bibfnamefont
  {N.}~\bibnamefont {Lavrentieva}}, \bibinfo {author} {\bibfnamefont
  {N.}~\bibnamefont {Lacome}}, \bibinfo {author} {\bibfnamefont {L.-H.}\
  \bibnamefont {Xu}}, \bibinfo {author} {\bibfnamefont {O.~M.}\ \bibnamefont
  {Lyulin}}, \bibinfo {author} {\bibfnamefont {J.~Y.}\ \bibnamefont {Mandin}},
  \bibinfo {author} {\bibfnamefont {A.}~\bibnamefont {Maki}}, \bibinfo {author}
  {\bibfnamefont {S.}~\bibnamefont {Mikhailenko}}, \bibinfo {author}
  {\bibfnamefont {C.~E.}\ \bibnamefont {Miller}}, \bibinfo {author}
  {\bibfnamefont {T.}~\bibnamefont {Mishina}}, \bibinfo {author} {\bibfnamefont
  {N.}~\bibnamefont {Moazzen-Ahmadi}}, \bibinfo {author} {\bibfnamefont
  {H.~S.~P.}\ \bibnamefont {M\"{u}ller}}, \bibinfo {author} {\bibfnamefont
  {A.}~\bibnamefont {Nikitin}}, \bibinfo {author} {\bibfnamefont
  {J.}~\bibnamefont {Orphal}}, \bibinfo {author} {\bibfnamefont
  {V.}~\bibnamefont {Perevalov}}, \bibinfo {author} {\bibfnamefont
  {A.}~\bibnamefont {Perrin}}, \bibinfo {author} {\bibfnamefont {D.~T.}\
  \bibnamefont {Petkie}}, \bibinfo {author} {\bibfnamefont {A.}~\bibnamefont
  {Predoi-Cross}}, \bibinfo {author} {\bibfnamefont {C.~P.}\ \bibnamefont
  {Rinsland}}, \bibinfo {author} {\bibfnamefont {J.~J.}\ \bibnamefont
  {Remedios}}, \bibinfo {author} {\bibfnamefont {M.}~\bibnamefont {Rotger}},
  \bibinfo {author} {\bibfnamefont {M.~A.~H.}\ \bibnamefont {Smith}}, \bibinfo
  {author} {\bibfnamefont {K.}~\bibnamefont {Sung}}, \bibinfo {author}
  {\bibfnamefont {S.}~\bibnamefont {Tashkun}}, \bibinfo {author} {\bibfnamefont
  {J.}~\bibnamefont {Tennyson}}, \bibinfo {author} {\bibfnamefont {R.~A.}\
  \bibnamefont {Toth}}, \bibinfo {author} {\bibfnamefont {A.~C.}\ \bibnamefont
  {Vandaele}}, \ and\ \bibinfo {author} {\bibfnamefont {J.}~\bibnamefont
  {Vander~Auwera}},\ }\href {\doibase 10.1016/j.jqsrt.2011.06.004} {\bibfield
  {journal} {\bibinfo  {journal} {J. Quant. Spectrosc. Radiat. Transf.}\
  }\textbf {\bibinfo {volume} {112}},\ \bibinfo {pages} {2395} (\bibinfo {year}
  {2011})}\BibitemShut {NoStop}%
\bibitem [{\citenamefont {Pickett}\ \emph {et~al.}(1998)\citenamefont
  {Pickett}, \citenamefont {Poynter}, \citenamefont {Cohen}, \citenamefont
  {Delitsky}, \citenamefont {Pearson},\ and\ \citenamefont {M\"{u}ller}}]{JPL}%
  \BibitemOpen
  \bibfield  {author} {\bibinfo {author} {\bibfnamefont {H.~M.}\ \bibnamefont
  {Pickett}}, \bibinfo {author} {\bibfnamefont {R.~L.}\ \bibnamefont
  {Poynter}}, \bibinfo {author} {\bibfnamefont {E.~A.}\ \bibnamefont {Cohen}},
  \bibinfo {author} {\bibfnamefont {M.~L.}\ \bibnamefont {Delitsky}}, \bibinfo
  {author} {\bibfnamefont {J.~C.}\ \bibnamefont {Pearson}}, \ and\ \bibinfo
  {author} {\bibfnamefont {H.~S.~P.}\ \bibnamefont {M\"{u}ller}},\ }\href
  {\doibase 10.1016/S0022-4073(98)00091-0} {\bibfield  {journal} {\bibinfo
  {journal} {J. Quant. Spectrosc. Radiat. Transf.}\ }\textbf {\bibinfo {volume}
  {60}},\ \bibinfo {pages} {883} (\bibinfo {year} {1998})}\BibitemShut
  {NoStop}%
\bibitem [{\citenamefont {M\"{u}ller}\ \emph {et~al.}(2005)\citenamefont
  {M\"{u}ller}, \citenamefont {Schl\"{o}der}, \citenamefont {Stutzki},\ and\
  \citenamefont {Winnewisser}}]{CDMS:2005}%
  \BibitemOpen
  \bibfield  {author} {\bibinfo {author} {\bibfnamefont {H.~S.~P.}\
  \bibnamefont {M\"{u}ller}}, \bibinfo {author} {\bibfnamefont
  {F.}~\bibnamefont {Schl\"{o}der}}, \bibinfo {author} {\bibfnamefont
  {J.}~\bibnamefont {Stutzki}}, \ and\ \bibinfo {author} {\bibfnamefont
  {G.}~\bibnamefont {Winnewisser}},\ }\href {\doibase
  10.1016/j.molstruc.2005.01.027} {\bibfield  {journal} {\bibinfo  {journal}
  {J. Mol. Struct.}\ }\textbf {\bibinfo {volume} {742}},\ \bibinfo {pages}
  {215} (\bibinfo {year} {2005})}\BibitemShut {NoStop}%
\bibitem [{\citenamefont {Sharpe}\ \emph {et~al.}(2004)\citenamefont {Sharpe},
  \citenamefont {Johnson}, \citenamefont {Sams}, \citenamefont {Chu},
  \citenamefont {Rhoderick},\ and\ \citenamefont {Johnson}}]{PNNL}%
  \BibitemOpen
  \bibfield  {author} {\bibinfo {author} {\bibfnamefont {S.~W.}\ \bibnamefont
  {Sharpe}}, \bibinfo {author} {\bibfnamefont {T.~J.}\ \bibnamefont {Johnson}},
  \bibinfo {author} {\bibfnamefont {R.~L.}\ \bibnamefont {Sams}}, \bibinfo
  {author} {\bibfnamefont {P.~M.}\ \bibnamefont {Chu}}, \bibinfo {author}
  {\bibfnamefont {G.~C.}\ \bibnamefont {Rhoderick}}, \ and\ \bibinfo {author}
  {\bibfnamefont {P.~A.}\ \bibnamefont {Johnson}},\ }\href {\doibase
  10.1366/0003702042641281} {\bibfield  {journal} {\bibinfo  {journal} {Appl.
  Spectrosc.}\ }\textbf {\bibinfo {volume} {58}},\ \bibinfo {pages} {1452}
  (\bibinfo {year} {2004})}\BibitemShut {NoStop}%
\bibitem [{\citenamefont {Wenger}\ and\ \citenamefont
  {Champion}(1998)}]{STDS:1998}%
  \BibitemOpen
  \bibfield  {author} {\bibinfo {author} {\bibfnamefont {C.}~\bibnamefont
  {Wenger}}\ and\ \bibinfo {author} {\bibfnamefont {J.~P.}\ \bibnamefont
  {Champion}},\ }\href {\doibase 10.1016/S0022-4073(97)00106-4} {\bibfield
  {journal} {\bibinfo  {journal} {J. Quant. Spectrosc. Radiat. Transf.}\
  }\textbf {\bibinfo {volume} {59}},\ \bibinfo {pages} {471} (\bibinfo {year}
  {1998})}\BibitemShut {NoStop}%
\bibitem [{\citenamefont {Cs{\'{a}}sz{\'{a}}r}, \citenamefont {Allen},\ and\
  \citenamefont {Schaefer~III}(1998)}]{Csaszar98}%
  \BibitemOpen
  \bibfield  {author} {\bibinfo {author} {\bibfnamefont {A.~G.}\ \bibnamefont
  {Cs{\'{a}}sz{\'{a}}r}}, \bibinfo {author} {\bibfnamefont {W.~D.}\
  \bibnamefont {Allen}}, \ and\ \bibinfo {author} {\bibfnamefont {H.~F.}\
  \bibnamefont {Schaefer~III}},\ }\href {\doibase 10.1063/1.476449} {\bibfield
  {journal} {\bibinfo  {journal} {J. Chem. Phys.}\ }\textbf {\bibinfo {volume}
  {108}},\ \bibinfo {pages} {9751} (\bibinfo {year} {1998})}\BibitemShut
  {NoStop}%
\bibitem [{\citenamefont {Adler}, \citenamefont {Knizia},\ and\ \citenamefont
  {Werner}(2007)}]{Adler07}%
  \BibitemOpen
  \bibfield  {author} {\bibinfo {author} {\bibfnamefont {T.~B.}\ \bibnamefont
  {Adler}}, \bibinfo {author} {\bibfnamefont {G.}~\bibnamefont {Knizia}}, \
  and\ \bibinfo {author} {\bibfnamefont {H.-J.}\ \bibnamefont {Werner}},\
  }\href {\doibase 10.1063/1.2817618} {\bibfield  {journal} {\bibinfo
  {journal} {J. Chem. Phys.}\ }\textbf {\bibinfo {volume} {127}},\ \bibinfo
  {pages} {221106} (\bibinfo {year} {2007})}\BibitemShut {NoStop}%
\bibitem [{\citenamefont {Peterson}, \citenamefont {Adler},\ and\ \citenamefont
  {Werner}(2008)}]{Peterson08}%
  \BibitemOpen
  \bibfield  {author} {\bibinfo {author} {\bibfnamefont {K.~A.}\ \bibnamefont
  {Peterson}}, \bibinfo {author} {\bibfnamefont {T.~B.}\ \bibnamefont {Adler}},
  \ and\ \bibinfo {author} {\bibfnamefont {H.-J.}\ \bibnamefont {Werner}},\
  }\href {\doibase 10.1063/1.2831537} {\bibfield  {journal} {\bibinfo
  {journal} {J. Chem. Phys.}\ }\textbf {\bibinfo {volume} {128}},\ \bibinfo
  {pages} {084102} (\bibinfo {year} {2008})}\BibitemShut {NoStop}%
\bibitem [{\citenamefont {Ten-No}(2004)}]{TenNo04}%
  \BibitemOpen
  \bibfield  {author} {\bibinfo {author} {\bibfnamefont {S.}~\bibnamefont
  {Ten-No}},\ }\href {\doibase 10.1016/j.cplett.2004.09.041} {\bibfield
  {journal} {\bibinfo  {journal} {Chem. Phys. Lett.}\ }\textbf {\bibinfo
  {volume} {398}},\ \bibinfo {pages} {56} (\bibinfo {year} {2004})}\BibitemShut
  {NoStop}%
\bibitem [{\citenamefont {Hill}\ \emph {et~al.}(2009)\citenamefont {Hill},
  \citenamefont {Peterson}, \citenamefont {Knizia},\ and\ \citenamefont
  {Werner}}]{Hill09}%
  \BibitemOpen
  \bibfield  {author} {\bibinfo {author} {\bibfnamefont {J.~G.}\ \bibnamefont
  {Hill}}, \bibinfo {author} {\bibfnamefont {K.~A.}\ \bibnamefont {Peterson}},
  \bibinfo {author} {\bibfnamefont {G.}~\bibnamefont {Knizia}}, \ and\ \bibinfo
  {author} {\bibfnamefont {H.-J.}\ \bibnamefont {Werner}},\ }\href {\doibase
  10.1063/1.3265857} {\bibfield  {journal} {\bibinfo  {journal} {J. Chem.
  Phys.}\ }\textbf {\bibinfo {volume} {131}},\ \bibinfo {pages} {194105}
  (\bibinfo {year} {2009})}\BibitemShut {NoStop}%
\bibitem [{\citenamefont {Yousaf}\ and\ \citenamefont
  {Peterson}(2008)}]{Yousaf08}%
  \BibitemOpen
  \bibfield  {author} {\bibinfo {author} {\bibfnamefont {K.~E.}\ \bibnamefont
  {Yousaf}}\ and\ \bibinfo {author} {\bibfnamefont {K.~A.}\ \bibnamefont
  {Peterson}},\ }\href {\doibase 10.1063/1.3009271} {\bibfield  {journal}
  {\bibinfo  {journal} {J. Chem. Phys.}\ }\textbf {\bibinfo {volume} {129}},\
  \bibinfo {pages} {184108} (\bibinfo {year} {2008})}\BibitemShut {NoStop}%
\bibitem [{\citenamefont {Weigend}(2002)}]{Weigend02}%
  \BibitemOpen
  \bibfield  {author} {\bibinfo {author} {\bibfnamefont {F.}~\bibnamefont
  {Weigend}},\ }\href {\doibase 10.1039/b204199p} {\bibfield  {journal}
  {\bibinfo  {journal} {Phys. Chem. Chem. Phys.}\ }\textbf {\bibinfo {volume}
  {4}},\ \bibinfo {pages} {4285} (\bibinfo {year} {2002})}\BibitemShut
  {NoStop}%
\bibitem [{\citenamefont {H{\"{a}}ttig}(2005)}]{Hattig05}%
  \BibitemOpen
  \bibfield  {author} {\bibinfo {author} {\bibfnamefont {C.}~\bibnamefont
  {H{\"{a}}ttig}},\ }\href {\doibase 10.1039/b415208e} {\bibfield  {journal}
  {\bibinfo  {journal} {Phys. Chem. Chem. Phys.}\ }\textbf {\bibinfo {volume}
  {7}},\ \bibinfo {pages} {59} (\bibinfo {year} {2005})}\BibitemShut {NoStop}%
\bibitem [{\citenamefont {Werner}\ \emph {et~al.}(2012)\citenamefont {Werner},
  \citenamefont {Knowles}, \citenamefont {Knizia}, \citenamefont {Manby},\ and\
  \citenamefont {Schuetz}}]{Werner2012}%
  \BibitemOpen
  \bibfield  {author} {\bibinfo {author} {\bibfnamefont {H.-J.}\ \bibnamefont
  {Werner}}, \bibinfo {author} {\bibfnamefont {P.~J.}\ \bibnamefont {Knowles}},
  \bibinfo {author} {\bibfnamefont {G.}~\bibnamefont {Knizia}}, \bibinfo
  {author} {\bibfnamefont {F.~R.}\ \bibnamefont {Manby}}, \ and\ \bibinfo
  {author} {\bibfnamefont {M.}~\bibnamefont {Schuetz}},\ }\href {\doibase
  10.1002/wcms.82} {\bibfield  {journal} {\bibinfo  {journal} {Comp. Mol.
  Sci.}\ }\textbf {\bibinfo {volume} {2}},\ \bibinfo {pages} {242} (\bibinfo
  {year} {2012})}\BibitemShut {NoStop}%
\bibitem [{\citenamefont {Douglas}\ and\ \citenamefont {Kroll}(1974)}]{dk1}%
  \BibitemOpen
  \bibfield  {author} {\bibinfo {author} {\bibfnamefont {M.}~\bibnamefont
  {Douglas}}\ and\ \bibinfo {author} {\bibfnamefont {N.~M.}\ \bibnamefont
  {Kroll}},\ }\href {\doibase 10.1016/0003-4916(74)90333-9} {\bibfield
  {journal} {\bibinfo  {journal} {Ann. Phys.}\ }\textbf {\bibinfo {volume}
  {82}},\ \bibinfo {pages} {89} (\bibinfo {year} {1974})}\BibitemShut {NoStop}%
\bibitem [{\citenamefont {He{\ss}}(1986)}]{dk2}%
  \BibitemOpen
  \bibfield  {author} {\bibinfo {author} {\bibfnamefont {B.~A.}\ \bibnamefont
  {He{\ss}}},\ }\href {\doibase 10.1103/PhysRevA.33.3742} {\bibfield  {journal}
  {\bibinfo  {journal} {Phys. Rev. A}\ }\textbf {\bibinfo {volume} {33}},\
  \bibinfo {pages} {3742} (\bibinfo {year} {1986})}\BibitemShut {NoStop}%
\bibitem [{\citenamefont {de~Jong}, \citenamefont {Harrison},\ and\
  \citenamefont {Dixon}(2001)}]{dk_basis}%
  \BibitemOpen
  \bibfield  {author} {\bibinfo {author} {\bibfnamefont {W.~A.}\ \bibnamefont
  {de~Jong}}, \bibinfo {author} {\bibfnamefont {R.~J.}\ \bibnamefont
  {Harrison}}, \ and\ \bibinfo {author} {\bibfnamefont {D.~A.}\ \bibnamefont
  {Dixon}},\ }\href {\doibase 10.1063/1.1329891} {\bibfield  {journal}
  {\bibinfo  {journal} {J. Chem. Phys.}\ }\textbf {\bibinfo {volume} {114}},\
  \bibinfo {pages} {48} (\bibinfo {year} {2001})}\BibitemShut {NoStop}%
\bibitem [{\citenamefont {Tarczay}\ \emph {et~al.}(2001)\citenamefont
  {Tarczay}, \citenamefont {Cs{\'{a}}sz{\'{a}}r}, \citenamefont {Klopper},\
  and\ \citenamefont {Quiney}}]{Tarczay01}%
  \BibitemOpen
  \bibfield  {author} {\bibinfo {author} {\bibfnamefont {G.}~\bibnamefont
  {Tarczay}}, \bibinfo {author} {\bibfnamefont {A.~G.}\ \bibnamefont
  {Cs{\'{a}}sz{\'{a}}r}}, \bibinfo {author} {\bibfnamefont {W.}~\bibnamefont
  {Klopper}}, \ and\ \bibinfo {author} {\bibfnamefont {H.~M.}\ \bibnamefont
  {Quiney}},\ }\href {\doibase 10.1080/00268970110073907} {\bibfield  {journal}
  {\bibinfo  {journal} {Mol. Phys.}\ }\textbf {\bibinfo {volume} {99}},\
  \bibinfo {pages} {1769} (\bibinfo {year} {2001})}\BibitemShut {NoStop}%
\bibitem [{\citenamefont {Hill}, \citenamefont {Mazumder},\ and\ \citenamefont
  {Peterson}(2010)}]{Hill10}%
  \BibitemOpen
  \bibfield  {author} {\bibinfo {author} {\bibfnamefont {J.~G.}\ \bibnamefont
  {Hill}}, \bibinfo {author} {\bibfnamefont {S.}~\bibnamefont {Mazumder}}, \
  and\ \bibinfo {author} {\bibfnamefont {K.~A.}\ \bibnamefont {Peterson}},\
  }\href {\doibase 10.1063/1.3308483} {\bibfield  {journal} {\bibinfo
  {journal} {J. Chem. Phys.}\ }\textbf {\bibinfo {volume} {132}},\ \bibinfo
  {pages} {054108} (\bibinfo {year} {2010})}\BibitemShut {NoStop}%
\bibitem [{\citenamefont {K\'{a}llay}\ and\ \citenamefont
  {Gauss}(2005)}]{Kallay05}%
  \BibitemOpen
  \bibfield  {author} {\bibinfo {author} {\bibfnamefont {M.}~\bibnamefont
  {K\'{a}llay}}\ and\ \bibinfo {author} {\bibfnamefont {J.}~\bibnamefont
  {Gauss}},\ }\href {\doibase 10.1063/1.2121589} {\bibfield  {journal}
  {\bibinfo  {journal} {J. Chem. Phys.}\ }\textbf {\bibinfo {volume} {123}},\
  \bibinfo {pages} {214105} (\bibinfo {year} {2005})}\BibitemShut {NoStop}%
\bibitem [{\citenamefont {K\'{a}llay}\ and\ \citenamefont
  {Gauss}(2008)}]{Kallay08}%
  \BibitemOpen
  \bibfield  {author} {\bibinfo {author} {\bibfnamefont {M.}~\bibnamefont
  {K\'{a}llay}}\ and\ \bibinfo {author} {\bibfnamefont {J.}~\bibnamefont
  {Gauss}},\ }\href {\doibase 10.1063/1.2988052} {\bibfield  {journal}
  {\bibinfo  {journal} {J. Chem. Phys.}\ }\textbf {\bibinfo {volume} {129}},\
  \bibinfo {pages} {144101} (\bibinfo {year} {2008})}\BibitemShut {NoStop}%
\bibitem [{MRCC()}]{mrcc}%
  \BibitemOpen
  MRCC,\ \href@noop {} {}\bibinfo {note} {A string-based quantum chemical
  program suite written by M. K{\'{a}}llay, See also M. K{\'{a}}llay and P. R.
  Surj{\'{a}}n, J. Chem. Phys. \textbf{115}, 2945 (2001) as well as
  www.mrcc.hu.}\BibitemShut {Stop}%
\bibitem [{CFOUR()}]{cfour}%
  \BibitemOpen
  CFOUR,\ \href@noop {} {}\bibinfo {note} {A quantum chemical program package
  written by J. F. Stanton, J. Gauss, M. E. Harding, and P. G. Szalay with
  contributions from A. A. Auer, R. J. Bartlett, U. Benedikt, C. Berger, D. E.
  Bernholdt, Y. J. Bomble, L. Cheng, O. Christiansen, M. Heckert, O. Heun, C.
  Huber, T.-C. Jagau, D. Jonsson, J. Jus{\'{e}}lius, K. Klein, W. J.
  Lauderdale, D. A. Matthews, T. Metzroth, L. A. M{\"{u}}ck, D. P. O'Neill, D.
  R. Price, E. Prochnow, C. Puzzarini, K. Ruud, F. Schiffmann, W. Schwalbach,
  S. Stopkowicz, A. Tajti, J. V{\'{a}}zquez, F. Wang, J. D. Watts, and the
  integral packages MOLECULE (J. Alml{\"{o}}f and P. R. Taylor), PROPS (P. R.
  Taylor), ABACUS (T. Helgaker, H. J. Aa. Jensen, P. J{\o}rgensen, and J.
  Olsen), and ECP routines by A. V. Mitin and C. van W{\"{u}}llen. For the
  current version, see http://www.cfour.de.}\BibitemShut {Stop}%
\bibitem [{\citenamefont {Kendall}, \citenamefont {Dunning~Jr.},\ and\
  \citenamefont {Harrison}(1992)}]{Kendall92}%
  \BibitemOpen
  \bibfield  {author} {\bibinfo {author} {\bibfnamefont {R.~A.}\ \bibnamefont
  {Kendall}}, \bibinfo {author} {\bibfnamefont {T.~H.}\ \bibnamefont
  {Dunning~Jr.}}, \ and\ \bibinfo {author} {\bibfnamefont {R.~J.}\ \bibnamefont
  {Harrison}},\ }\href {\doibase 10.1063/1.462569} {\bibfield  {journal}
  {\bibinfo  {journal} {J. Chem. Phys.}\ }\textbf {\bibinfo {volume} {96}},\
  \bibinfo {pages} {6796} (\bibinfo {year} {1992})}\BibitemShut {NoStop}%
\bibitem [{\citenamefont {Woon}\ and\ \citenamefont
  {Dunning~Jr.}(1993)}]{Woon93}%
  \BibitemOpen
  \bibfield  {author} {\bibinfo {author} {\bibfnamefont {D.~E.}\ \bibnamefont
  {Woon}}\ and\ \bibinfo {author} {\bibfnamefont {T.~H.}\ \bibnamefont
  {Dunning~Jr.}},\ }\href {\doibase 10.1063/1.464303} {\bibfield  {journal}
  {\bibinfo  {journal} {J. Chem. Phys.}\ }\textbf {\bibinfo {volume} {98}},\
  \bibinfo {pages} {1358} (\bibinfo {year} {1993})}\BibitemShut {NoStop}%
\bibitem [{\citenamefont {Dunning~Jr.}, \citenamefont {Peterson},\ and\
  \citenamefont {Wilson}(2001)}]{Dunning01}%
  \BibitemOpen
  \bibfield  {author} {\bibinfo {author} {\bibfnamefont {T.~H.}\ \bibnamefont
  {Dunning~Jr.}}, \bibinfo {author} {\bibfnamefont {K.~A.}\ \bibnamefont
  {Peterson}}, \ and\ \bibinfo {author} {\bibfnamefont {A.~K.}\ \bibnamefont
  {Wilson}},\ }\href {\doibase 10.1063/1.1367373} {\bibfield  {journal}
  {\bibinfo  {journal} {J. Chem. Phys.}\ }\textbf {\bibinfo {volume} {114}},\
  \bibinfo {pages} {9244} (\bibinfo {year} {2001})}\BibitemShut {NoStop}%
\bibitem [{\citenamefont {Gauss}\ \emph {et~al.}(2006)\citenamefont {Gauss},
  \citenamefont {Tajti}, \citenamefont {K{\'{a}}llay}, \citenamefont
  {Stanton},\ and\ \citenamefont {Szalay}}]{Gauss06}%
  \BibitemOpen
  \bibfield  {author} {\bibinfo {author} {\bibfnamefont {J.}~\bibnamefont
  {Gauss}}, \bibinfo {author} {\bibfnamefont {A.}~\bibnamefont {Tajti}},
  \bibinfo {author} {\bibfnamefont {M.}~\bibnamefont {K{\'{a}}llay}}, \bibinfo
  {author} {\bibfnamefont {J.~F.}\ \bibnamefont {Stanton}}, \ and\ \bibinfo
  {author} {\bibfnamefont {P.~G.}\ \bibnamefont {Szalay}},\ }\href {\doibase
  10.1063/1.2356465} {\bibfield  {journal} {\bibinfo  {journal} {J. Chem.
  Phys.}\ }\textbf {\bibinfo {volume} {125}},\ \bibinfo {pages} {144111}
  (\bibinfo {year} {2006})}\BibitemShut {NoStop}%
\bibitem [{\citenamefont {Yachmenev}\ \emph {et~al.}(2011)\citenamefont
  {Yachmenev}, \citenamefont {Yurchenko}, \citenamefont {Ribeyre},\ and\
  \citenamefont {Thiel}}]{YaYuRi11.H2CS}%
  \BibitemOpen
  \bibfield  {author} {\bibinfo {author} {\bibfnamefont {A.}~\bibnamefont
  {Yachmenev}}, \bibinfo {author} {\bibfnamefont {S.~N.}\ \bibnamefont
  {Yurchenko}}, \bibinfo {author} {\bibfnamefont {T.}~\bibnamefont {Ribeyre}},
  \ and\ \bibinfo {author} {\bibfnamefont {W.}~\bibnamefont {Thiel}},\ }\href
  {\doibase 10.1063/1.3624570} {\bibfield  {journal} {\bibinfo  {journal} {J.
  Chem. Phys.}\ }\textbf {\bibinfo {volume} {135}},\ \bibinfo {pages} {074302}
  (\bibinfo {year} {2011})}\BibitemShut {NoStop}%
\bibitem [{\citenamefont {Owens}\ \emph {et~al.}(2015)\citenamefont {Owens},
  \citenamefont {Yurchenko}, \citenamefont {Yachmenev}, \citenamefont
  {Tennyson},\ and\ \citenamefont {Thiel}}]{15OwYuYa.CH3Cl}%
  \BibitemOpen
  \bibfield  {author} {\bibinfo {author} {\bibfnamefont {A.}~\bibnamefont
  {Owens}}, \bibinfo {author} {\bibfnamefont {S.~N.}\ \bibnamefont
  {Yurchenko}}, \bibinfo {author} {\bibfnamefont {A.}~\bibnamefont
  {Yachmenev}}, \bibinfo {author} {\bibfnamefont {J.}~\bibnamefont {Tennyson}},
  \ and\ \bibinfo {author} {\bibfnamefont {W.}~\bibnamefont {Thiel}},\ }\href
  {\doibase 10.1063/1.4922890} {\bibfield  {journal} {\bibinfo  {journal} {J.
  Chem. Phys.}\ }\textbf {\bibinfo {volume} {142}},\ \bibinfo {pages} {244306}
  (\bibinfo {year} {2015})}\BibitemShut {NoStop}%
\bibitem [{\citenamefont {Yurchenko}\ \emph {et~al.}(2013)\citenamefont
  {Yurchenko}, \citenamefont {Tennyson}, \citenamefont {Barber},\ and\
  \citenamefont {Thiel}}]{13YuTeBa.CH4}%
  \BibitemOpen
  \bibfield  {author} {\bibinfo {author} {\bibfnamefont {S.~N.}\ \bibnamefont
  {Yurchenko}}, \bibinfo {author} {\bibfnamefont {J.}~\bibnamefont {Tennyson}},
  \bibinfo {author} {\bibfnamefont {R.~J.}\ \bibnamefont {Barber}}, \ and\
  \bibinfo {author} {\bibfnamefont {W.}~\bibnamefont {Thiel}},\ }\href
  {\doibase 10.1016/j.jms.2013.05.014} {\bibfield  {journal} {\bibinfo
  {journal} {J. Mol. Spectrosc.}\ }\textbf {\bibinfo {volume} {291}},\ \bibinfo
  {pages} {69} (\bibinfo {year} {2013})}\BibitemShut {NoStop}%
\bibitem [{\citenamefont {Yurchenko}\ and\ \citenamefont
  {Tennyson}(2014)}]{14YuTexx.CH4}%
  \BibitemOpen
  \bibfield  {author} {\bibinfo {author} {\bibfnamefont {S.~N.}\ \bibnamefont
  {Yurchenko}}\ and\ \bibinfo {author} {\bibfnamefont {J.}~\bibnamefont
  {Tennyson}},\ }\href {\doibase 10.1093/mnras/stu326} {\bibfield  {journal}
  {\bibinfo  {journal} {Mon. Not. R. Astron. Soc.}\ }\textbf {\bibinfo {volume}
  {440}},\ \bibinfo {pages} {1649} (\bibinfo {year} {2014})}\BibitemShut
  {NoStop}%
\bibitem [{\citenamefont {Yurchenko}\ \emph {et~al.}(2014)\citenamefont
  {Yurchenko}, \citenamefont {Tennyson}, \citenamefont {Bailey}, \citenamefont
  {Hollis},\ and\ \citenamefont {Tinetti}}]{14YuTeBa.CH4}%
  \BibitemOpen
  \bibfield  {author} {\bibinfo {author} {\bibfnamefont {S.~N.}\ \bibnamefont
  {Yurchenko}}, \bibinfo {author} {\bibfnamefont {J.}~\bibnamefont {Tennyson}},
  \bibinfo {author} {\bibfnamefont {J.}~\bibnamefont {Bailey}}, \bibinfo
  {author} {\bibfnamefont {M.~D.~J.}\ \bibnamefont {Hollis}}, \ and\ \bibinfo
  {author} {\bibfnamefont {G.}~\bibnamefont {Tinetti}},\ }\href {\doibase
  10.1073/pnas.1324219111} {\bibfield  {journal} {\bibinfo  {journal} {Proc.
  Natl. Acad. Sci. U. S. A.}\ }\textbf {\bibinfo {volume} {111}},\ \bibinfo
  {pages} {9379} (\bibinfo {year} {2014})}\BibitemShut {NoStop}%
\bibitem [{\citenamefont {Bunker}\ and\ \citenamefont
  {Jensen}(1998)}]{MolSym_BuJe98}%
  \BibitemOpen
  \bibfield  {author} {\bibinfo {author} {\bibfnamefont {P.~R.}\ \bibnamefont
  {Bunker}}\ and\ \bibinfo {author} {\bibfnamefont {P.}~\bibnamefont
  {Jensen}},\ }\href@noop {} {\emph {\bibinfo {title} {Molecular Symmetry and
  Spectroscopy}}},\ \bibinfo {edition} {2nd}\ ed.\ (\bibinfo  {publisher} {NRC
  Research Press, Ottawa},\ \bibinfo {year} {1998})\BibitemShut {NoStop}%
\bibitem [{\citenamefont {Partridge}\ and\ \citenamefont
  {Schwenke}(1997)}]{Schwenke97}%
  \BibitemOpen
  \bibfield  {author} {\bibinfo {author} {\bibfnamefont {H.}~\bibnamefont
  {Partridge}}\ and\ \bibinfo {author} {\bibfnamefont {D.~W.}\ \bibnamefont
  {Schwenke}},\ }\href {\doibase 10.1063/1.473987} {\bibfield  {journal}
  {\bibinfo  {journal} {J. Chem. Phys.}\ }\textbf {\bibinfo {volume} {106}},\
  \bibinfo {pages} {4618} (\bibinfo {year} {1997})}\BibitemShut {NoStop}%
\bibitem [{\citenamefont {Watson}(2003)}]{Watson03}%
  \BibitemOpen
  \bibfield  {author} {\bibinfo {author} {\bibfnamefont {J.~K.~G.}\
  \bibnamefont {Watson}},\ }\href {\doibase 10.1016/S0022-2852(03)00100-0}
  {\bibfield  {journal} {\bibinfo  {journal} {J. Mol. Spectrosc.}\ }\textbf
  {\bibinfo {volume} {219}},\ \bibinfo {pages} {326} (\bibinfo {year}
  {2003})}\BibitemShut {NoStop}%
\bibitem [{\citenamefont {Lee}\ and\ \citenamefont {Taylor}(1989)}]{T1_Lee89}%
  \BibitemOpen
  \bibfield  {author} {\bibinfo {author} {\bibfnamefont {T.~J.}\ \bibnamefont
  {Lee}}\ and\ \bibinfo {author} {\bibfnamefont {P.~R.}\ \bibnamefont
  {Taylor}},\ }\href {\doibase 10.1002/qua.560360824} {\bibfield  {journal}
  {\bibinfo  {journal} {Int. J. Quantum Chem.}\ }\textbf {\bibinfo {volume}
  {36}},\ \bibinfo {pages} {199} (\bibinfo {year} {1989})}\BibitemShut
  {NoStop}%
\bibitem [{EPA()}]{EPAPSSIH4}%
  \BibitemOpen
  \href@noop {} {}\bibinfo {note} {See supplementary material at ``insert URL''
  for the parameters of the potential energy surface and dipole moment surface
  for SiH$_4$.}\BibitemShut {Stop}%
\bibitem [{\citenamefont {Yurchenko}\ \emph {et~al.}(2009)\citenamefont
  {Yurchenko}, \citenamefont {Barber}, \citenamefont {Yachmenev}, \citenamefont
  {Thiel}, \citenamefont {Jensen},\ and\ \citenamefont
  {Tennyson}}]{YuBaYa09.NH3}%
  \BibitemOpen
  \bibfield  {author} {\bibinfo {author} {\bibfnamefont {S.~N.}\ \bibnamefont
  {Yurchenko}}, \bibinfo {author} {\bibfnamefont {R.~J.}\ \bibnamefont
  {Barber}}, \bibinfo {author} {\bibfnamefont {A.}~\bibnamefont {Yachmenev}},
  \bibinfo {author} {\bibfnamefont {W.}~\bibnamefont {Thiel}}, \bibinfo
  {author} {\bibfnamefont {P.}~\bibnamefont {Jensen}}, \ and\ \bibinfo {author}
  {\bibfnamefont {J.}~\bibnamefont {Tennyson}},\ }\href {\doibase
  10.1021/jp9029425} {\bibfield  {journal} {\bibinfo  {journal} {J. Phys. Chem.
  A}\ }\textbf {\bibinfo {volume} {113}},\ \bibinfo {pages} {11845} (\bibinfo
  {year} {2009})}\BibitemShut {NoStop}%
\bibitem [{\citenamefont {Yachmenev}, \citenamefont {Polyak},\ and\
  \citenamefont {Thiel}(2013)}]{13YaPoTh.H2CS}%
  \BibitemOpen
  \bibfield  {author} {\bibinfo {author} {\bibfnamefont {A.}~\bibnamefont
  {Yachmenev}}, \bibinfo {author} {\bibfnamefont {I.}~\bibnamefont {Polyak}}, \
  and\ \bibinfo {author} {\bibfnamefont {W.}~\bibnamefont {Thiel}},\ }\href
  {\doibase 10.1063/1.4832322} {\bibfield  {journal} {\bibinfo  {journal} {J.
  Chem. Phys.}\ }\textbf {\bibinfo {volume} {139}},\ \bibinfo {pages} {204308}
  (\bibinfo {year} {2013})}\BibitemShut {NoStop}%
\bibitem [{\citenamefont {Ohno}\ \emph {et~al.}(1985)\citenamefont {Ohno},
  \citenamefont {Matsuura}, \citenamefont {Endo},\ and\ \citenamefont
  {Hirota}}]{85OhMaEn.SiH4}%
  \BibitemOpen
  \bibfield  {author} {\bibinfo {author} {\bibfnamefont {K.}~\bibnamefont
  {Ohno}}, \bibinfo {author} {\bibfnamefont {H.}~\bibnamefont {Matsuura}},
  \bibinfo {author} {\bibfnamefont {Y.}~\bibnamefont {Endo}}, \ and\ \bibinfo
  {author} {\bibfnamefont {E.}~\bibnamefont {Hirota}},\ }\href {\doibase
  10.1016/0022-2852(85)90070-0} {\bibfield  {journal} {\bibinfo  {journal} {J.
  Mol. Spectrosc.}\ }\textbf {\bibinfo {volume} {111}},\ \bibinfo {pages} {73}
  (\bibinfo {year} {1985})}\BibitemShut {NoStop}%
\bibitem [{\citenamefont {Coriani}\ \emph {et~al.}(2005)\citenamefont
  {Coriani}, \citenamefont {Marchesan}, \citenamefont {Gauss}, \citenamefont
  {H\"{a}ttig}, \citenamefont {Helgaker},\ and\ \citenamefont
  {J{\o}rgensen}}]{05CoMaGa.SiH4}%
  \BibitemOpen
  \bibfield  {author} {\bibinfo {author} {\bibfnamefont {S.}~\bibnamefont
  {Coriani}}, \bibinfo {author} {\bibfnamefont {D.}~\bibnamefont {Marchesan}},
  \bibinfo {author} {\bibfnamefont {J.}~\bibnamefont {Gauss}}, \bibinfo
  {author} {\bibfnamefont {C.}~\bibnamefont {H\"{a}ttig}}, \bibinfo {author}
  {\bibfnamefont {T.}~\bibnamefont {Helgaker}}, \ and\ \bibinfo {author}
  {\bibfnamefont {P.}~\bibnamefont {J{\o}rgensen}},\ }\href {\doibase
  10.1063/1.2104387} {\bibfield  {journal} {\bibinfo  {journal} {J. Chem.
  Phys.}\ }\textbf {\bibinfo {volume} {123}},\ \bibinfo {pages} {184107}
  (\bibinfo {year} {2005})}\BibitemShut {NoStop}%
\bibitem [{\citenamefont {Yurchenko}, \citenamefont {Thiel},\ and\
  \citenamefont {Jensen}(2007)}]{TROVE2007}%
  \BibitemOpen
  \bibfield  {author} {\bibinfo {author} {\bibfnamefont {S.~N.}\ \bibnamefont
  {Yurchenko}}, \bibinfo {author} {\bibfnamefont {W.}~\bibnamefont {Thiel}}, \
  and\ \bibinfo {author} {\bibfnamefont {P.}~\bibnamefont {Jensen}},\ }\href
  {\doibase 10.1016/j.jms.2007.07.009} {\bibfield  {journal} {\bibinfo
  {journal} {J. Mol. Spectrosc.}\ }\textbf {\bibinfo {volume} {245}},\ \bibinfo
  {pages} {126} (\bibinfo {year} {2007})}\BibitemShut {NoStop}%
\bibitem [{\citenamefont {Yachmenev}\ and\ \citenamefont
  {Yurchenko}(2015)}]{15YaYu.ADF}%
  \BibitemOpen
  \bibfield  {author} {\bibinfo {author} {\bibfnamefont {A.}~\bibnamefont
  {Yachmenev}}\ and\ \bibinfo {author} {\bibfnamefont {S.~N.}\ \bibnamefont
  {Yurchenko}},\ }\href {\doibase 10.1063/1.4923039} {\bibfield  {journal}
  {\bibinfo  {journal} {J. Chem. Phys.}\ }\textbf {\bibinfo {volume} {143}},\
  \bibinfo {pages} {014105} (\bibinfo {year} {2015})}\BibitemShut {NoStop}%
\bibitem [{\citenamefont {Noumerov}(1924)}]{Numerov1924}%
  \BibitemOpen
  \bibfield  {author} {\bibinfo {author} {\bibfnamefont {B.~V.}\ \bibnamefont
  {Noumerov}},\ }\href@noop {} {\bibfield  {journal} {\bibinfo  {journal} {Mon.
  Not. R. Astron. Soc.}\ }\textbf {\bibinfo {volume} {84}},\ \bibinfo {pages}
  {592} (\bibinfo {year} {1924})}\BibitemShut {NoStop}%
\bibitem [{\citenamefont {Cooley}(1961)}]{Cooley1961}%
  \BibitemOpen
  \bibfield  {author} {\bibinfo {author} {\bibfnamefont {J.~W.}\ \bibnamefont
  {Cooley}},\ }\href@noop {} {\bibfield  {journal} {\bibinfo  {journal} {Math.
  Comput.}\ }\textbf {\bibinfo {volume} {15}},\ \bibinfo {pages} {363}
  (\bibinfo {year} {1961})}\BibitemShut {NoStop}%
\bibitem [{\citenamefont {Ovsyannikov}\ \emph {et~al.}(2008)\citenamefont
  {Ovsyannikov}, \citenamefont {Thiel}, \citenamefont {Yurchenko},
  \citenamefont {Carvajal},\ and\ \citenamefont {Jensen}}]{OvThYu08.PH3}%
  \BibitemOpen
  \bibfield  {author} {\bibinfo {author} {\bibfnamefont {R.~I.}\ \bibnamefont
  {Ovsyannikov}}, \bibinfo {author} {\bibfnamefont {W.}~\bibnamefont {Thiel}},
  \bibinfo {author} {\bibfnamefont {S.~N.}\ \bibnamefont {Yurchenko}}, \bibinfo
  {author} {\bibfnamefont {M.}~\bibnamefont {Carvajal}}, \ and\ \bibinfo
  {author} {\bibfnamefont {P.}~\bibnamefont {Jensen}},\ }\href {\doibase
  10.1063/1.2956488} {\bibfield  {journal} {\bibinfo  {journal} {J. Chem.
  Phys.}\ }\textbf {\bibinfo {volume} {129}},\ \bibinfo {pages} {044309}
  (\bibinfo {year} {2008})}\BibitemShut {NoStop}%
\bibitem [{\citenamefont {Chevalier}()}]{Chevalier:1988}%
  \BibitemOpen
  \bibfield  {author} {\bibinfo {author} {\bibfnamefont {M.}~\bibnamefont
  {Chevalier}},\ }\href@noop {} {}\bibinfo {note} {{Thesis, Universit\'{e} de
  Paris Sud, France, (1988)}}\BibitemShut {NoStop}%
\bibitem [{\citenamefont {Bernheim}\ \emph {et~al.}(1984)\citenamefont
  {Bernheim}, \citenamefont {Lampe}, \citenamefont {O'Keefe},\ and\
  \citenamefont {Qualey~{III}}}]{84BeLaOk.SiH4}%
  \BibitemOpen
  \bibfield  {author} {\bibinfo {author} {\bibfnamefont {R.~A.}\ \bibnamefont
  {Bernheim}}, \bibinfo {author} {\bibfnamefont {F.~W.}\ \bibnamefont {Lampe}},
  \bibinfo {author} {\bibfnamefont {J.~F.}\ \bibnamefont {O'Keefe}}, \ and\
  \bibinfo {author} {\bibfnamefont {J.~R.}\ \bibnamefont {Qualey~{III}}},\
  }\href {\doibase 10.1063/1.446694} {\bibfield  {journal} {\bibinfo  {journal}
  {J. Chem. Phys.}\ }\textbf {\bibinfo {volume} {80}},\ \bibinfo {pages} {5906}
  (\bibinfo {year} {1984})}\BibitemShut {NoStop}%
\bibitem [{\citenamefont {Fox}\ and\ \citenamefont
  {Person}(1976)}]{76FoPexx.SiH4}%
  \BibitemOpen
  \bibfield  {author} {\bibinfo {author} {\bibfnamefont {K.}~\bibnamefont
  {Fox}}\ and\ \bibinfo {author} {\bibfnamefont {W.~B.}\ \bibnamefont
  {Person}},\ }\href {\doibase 10.1063/1.432196} {\bibfield  {journal}
  {\bibinfo  {journal} {J. Chem. Phys.}\ }\textbf {\bibinfo {volume} {64}},\
  \bibinfo {pages} {5218} (\bibinfo {year} {1976})}\BibitemShut {NoStop}%
\bibitem [{\citenamefont {Ball}\ and\ \citenamefont
  {McKean}(1962)}]{62BaMcxx.SiH4}%
  \BibitemOpen
  \bibfield  {author} {\bibinfo {author} {\bibfnamefont {D.~F.}\ \bibnamefont
  {Ball}}\ and\ \bibinfo {author} {\bibfnamefont {D.~C.}\ \bibnamefont
  {McKean}},\ }\href {\doibase 10.1016/S0371-1951(62)80217-3} {\bibfield
  {journal} {\bibinfo  {journal} {Spectroc. Acta}\ }\textbf {\bibinfo {volume}
  {18}},\ \bibinfo {pages} {1019} (\bibinfo {year} {1962})}\BibitemShut
  {NoStop}%
\bibitem [{\citenamefont {Levin}\ and\ \citenamefont
  {King}(1962)}]{62LeKixx.SiH4}%
  \BibitemOpen
  \bibfield  {author} {\bibinfo {author} {\bibfnamefont {I.~W.}\ \bibnamefont
  {Levin}}\ and\ \bibinfo {author} {\bibfnamefont {W.~T.}\ \bibnamefont
  {King}},\ }\href {\doibase 10.1063/1.1733292} {\bibfield  {journal} {\bibinfo
   {journal} {J. Chem. Phys.}\ }\textbf {\bibinfo {volume} {37}},\ \bibinfo
  {pages} {1375} (\bibinfo {year} {1962})}\BibitemShut {NoStop}%
\bibitem [{\citenamefont {Coats}, \citenamefont {McKean},\ and\ \citenamefont
  {Steele}(1994)}]{94CoMcSt.SiH4}%
  \BibitemOpen
  \bibfield  {author} {\bibinfo {author} {\bibfnamefont {A.~M.}\ \bibnamefont
  {Coats}}, \bibinfo {author} {\bibfnamefont {D.~C.}\ \bibnamefont {McKean}}, \
  and\ \bibinfo {author} {\bibfnamefont {D.}~\bibnamefont {Steele}},\ }\href
  {\doibase 10.1016/0022-2860(94)07974-9} {\bibfield  {journal} {\bibinfo
  {journal} {J. Mol. Struct.}\ }\textbf {\bibinfo {volume} {320}},\ \bibinfo
  {pages} {269} (\bibinfo {year} {1994})}\BibitemShut {NoStop}%
\bibitem [{\citenamefont {Cadot}(1992)}]{92Caxxxx.SiH4}%
  \BibitemOpen
  \bibfield  {author} {\bibinfo {author} {\bibfnamefont {J.}~\bibnamefont
  {Cadot}},\ }\href {\doibase 10.1016/0022-2852(92)90216-B} {\bibfield
  {journal} {\bibinfo  {journal} {J. Mol. Spectrosc.}\ }\textbf {\bibinfo
  {volume} {154}},\ \bibinfo {pages} {383} (\bibinfo {year}
  {1992})}\BibitemShut {NoStop}%
\bibitem [{\citenamefont {Fox}(1970)}]{Fox:1970}%
  \BibitemOpen
  \bibfield  {author} {\bibinfo {author} {\bibfnamefont {K.}~\bibnamefont
  {Fox}},\ }\href {\doibase 10.1016/0022-4073(70)90015-4} {\bibfield  {journal}
  {\bibinfo  {journal} {J. Quant. Spectrosc. Radiat. Transf.}\ }\textbf
  {\bibinfo {volume} {10}},\ \bibinfo {pages} {1335} (\bibinfo {year}
  {1970})}\BibitemShut {NoStop}%
\bibitem [{\citenamefont {Dang-Nhu}, \citenamefont {Pierre},\ and\
  \citenamefont {Saint-Loup}(1974)}]{74DaPiSa.SiH4}%
  \BibitemOpen
  \bibfield  {author} {\bibinfo {author} {\bibfnamefont {M.}~\bibnamefont
  {Dang-Nhu}}, \bibinfo {author} {\bibfnamefont {G.}~\bibnamefont {Pierre}}, \
  and\ \bibinfo {author} {\bibfnamefont {R.}~\bibnamefont {Saint-Loup}},\
  }\href {\doibase 10.1080/00268977400102981} {\bibfield  {journal} {\bibinfo
  {journal} {Mol. Phys.}\ }\textbf {\bibinfo {volume} {28}},\ \bibinfo {pages}
  {447} (\bibinfo {year} {1974})}\BibitemShut {NoStop}%
\bibitem [{\citenamefont {Pierre}, \citenamefont {Guelachvili},\ and\
  \citenamefont {Amiot}(1975)}]{75PiGuAm.SiH4}%
  \BibitemOpen
  \bibfield  {author} {\bibinfo {author} {\bibfnamefont {G.}~\bibnamefont
  {Pierre}}, \bibinfo {author} {\bibfnamefont {G.}~\bibnamefont {Guelachvili}},
  \ and\ \bibinfo {author} {\bibfnamefont {C.}~\bibnamefont {Amiot}},\ }\href
  {\doibase 10.1051/jphys:01975003606048700} {\bibfield  {journal} {\bibinfo
  {journal} {J. Phys. France}\ }\textbf {\bibinfo {volume} {36}},\ \bibinfo
  {pages} {487} (\bibinfo {year} {1975})}\BibitemShut {NoStop}%
\bibitem [{\citenamefont {Pierre}, \citenamefont {Valentin},\ and\
  \citenamefont {Henry}(1984)}]{84PiVaHe.SiH4}%
  \BibitemOpen
  \bibfield  {author} {\bibinfo {author} {\bibfnamefont {G.}~\bibnamefont
  {Pierre}}, \bibinfo {author} {\bibfnamefont {A.}~\bibnamefont {Valentin}}, \
  and\ \bibinfo {author} {\bibfnamefont {L.}~\bibnamefont {Henry}},\
  }\href@noop {} {\bibfield  {journal} {\bibinfo  {journal} {Can. J. Phys.}\
  }\textbf {\bibinfo {volume} {62}},\ \bibinfo {pages} {254} (\bibinfo {year}
  {1984})}\BibitemShut {NoStop}%
\bibitem [{\citenamefont {van Helden}\ \emph {et~al.}(2015)\citenamefont {van
  Helden}, \citenamefont {Lopatik}, \citenamefont {Nave}, \citenamefont {Lang},
  \citenamefont {Davies},\ and\ \citenamefont {R\"{o}pcke}}]{15HeLoNa.SiH4}%
  \BibitemOpen
  \bibfield  {author} {\bibinfo {author} {\bibfnamefont {J.~H.}\ \bibnamefont
  {van Helden}}, \bibinfo {author} {\bibfnamefont {D.}~\bibnamefont {Lopatik}},
  \bibinfo {author} {\bibfnamefont {A.}~\bibnamefont {Nave}}, \bibinfo {author}
  {\bibfnamefont {N.}~\bibnamefont {Lang}}, \bibinfo {author} {\bibfnamefont
  {P.~B.}\ \bibnamefont {Davies}}, \ and\ \bibinfo {author} {\bibfnamefont
  {J.}~\bibnamefont {R\"{o}pcke}},\ }\href {\doibase
  10.1016/j.jqsrt.2014.10.016} {\bibfield  {journal} {\bibinfo  {journal} {J.
  Quant. Spectrosc. Radiat. Transf.}\ }\textbf {\bibinfo {volume} {151}},\
  \bibinfo {pages} {287} (\bibinfo {year} {2015})}\BibitemShut {NoStop}%
\bibitem [{\citenamefont {Yurchenko}\ \emph {et~al.}(2011)\citenamefont
  {Yurchenko}, \citenamefont {Barber}, \citenamefont {Tennyson}, \citenamefont
  {Thiel},\ and\ \citenamefont {Jensen}}]{YuBaTe11.NH3}%
  \BibitemOpen
  \bibfield  {author} {\bibinfo {author} {\bibfnamefont {S.~N.}\ \bibnamefont
  {Yurchenko}}, \bibinfo {author} {\bibfnamefont {R.~J.}\ \bibnamefont
  {Barber}}, \bibinfo {author} {\bibfnamefont {J.}~\bibnamefont {Tennyson}},
  \bibinfo {author} {\bibfnamefont {W.}~\bibnamefont {Thiel}}, \ and\ \bibinfo
  {author} {\bibfnamefont {P.}~\bibnamefont {Jensen}},\ }\href {\doibase
  10.1016/j.jms.2011.04.005} {\bibfield  {journal} {\bibinfo  {journal} {J.
  Mol. Spectrosc.}\ }\textbf {\bibinfo {volume} {268}},\ \bibinfo {pages} {123}
  (\bibinfo {year} {2011})}\BibitemShut {NoStop}%
\bibitem [{\citenamefont {Tennyson}\ \emph {et~al.}(2016)\citenamefont
  {Tennyson}, \citenamefont {Yurchenko}, \citenamefont {Al-Refaie},
  \citenamefont {Barton}, \citenamefont {Chubb}, \citenamefont {Coles},
  \citenamefont {Diamantopoulou}, \citenamefont {Gorman}, \citenamefont {Hill},
  \citenamefont {Lam}, \citenamefont {Lodi}, \citenamefont {McKemmish},
  \citenamefont {Na}, \citenamefont {Owens}, \citenamefont {Polyansky},
  \citenamefont {Rivlin}, \citenamefont {Sousa-Silva}, \citenamefont
  {Underwood}, \citenamefont {Yachmenev},\ and\ \citenamefont
  {Zak}}]{ExoMol2016}%
  \BibitemOpen
  \bibfield  {author} {\bibinfo {author} {\bibfnamefont {J.}~\bibnamefont
  {Tennyson}}, \bibinfo {author} {\bibfnamefont {S.~N.}\ \bibnamefont
  {Yurchenko}}, \bibinfo {author} {\bibfnamefont {A.~F.}\ \bibnamefont
  {Al-Refaie}}, \bibinfo {author} {\bibfnamefont {E.~J.}\ \bibnamefont
  {Barton}}, \bibinfo {author} {\bibfnamefont {K.~L.}\ \bibnamefont {Chubb}},
  \bibinfo {author} {\bibfnamefont {P.~A.}\ \bibnamefont {Coles}}, \bibinfo
  {author} {\bibfnamefont {S.}~\bibnamefont {Diamantopoulou}}, \bibinfo
  {author} {\bibfnamefont {M.~N.}\ \bibnamefont {Gorman}}, \bibinfo {author}
  {\bibfnamefont {C.}~\bibnamefont {Hill}}, \bibinfo {author} {\bibfnamefont
  {A.~Z.}\ \bibnamefont {Lam}}, \bibinfo {author} {\bibfnamefont
  {L.}~\bibnamefont {Lodi}}, \bibinfo {author} {\bibfnamefont {L.~K.}\
  \bibnamefont {McKemmish}}, \bibinfo {author} {\bibfnamefont {Y.}~\bibnamefont
  {Na}}, \bibinfo {author} {\bibfnamefont {A.}~\bibnamefont {Owens}}, \bibinfo
  {author} {\bibfnamefont {O.~L.}\ \bibnamefont {Polyansky}}, \bibinfo {author}
  {\bibfnamefont {T.}~\bibnamefont {Rivlin}}, \bibinfo {author} {\bibfnamefont
  {C.}~\bibnamefont {Sousa-Silva}}, \bibinfo {author} {\bibfnamefont {D.~S.}\
  \bibnamefont {Underwood}}, \bibinfo {author} {\bibfnamefont {A.}~\bibnamefont
  {Yachmenev}}, \ and\ \bibinfo {author} {\bibfnamefont {E.}~\bibnamefont
  {Zak}},\ }\href@noop {} {\bibfield  {journal} {\bibinfo  {journal} {J. Mol.
  Spectrosc.}\ }\textbf {\bibinfo {volume} {327}},\ \bibinfo {pages} {73}
  (\bibinfo {year} {2016})}\BibitemShut {NoStop}%
\end{thebibliography}
%

\end{document}